\documentclass[aps,prb,twocolumn,groupedaddress,floatfix]{revtex4}

\usepackage [dvips]{graphicx}

\begin{document}
%\preprint{}
\title{Exploring the high pressure phase diagram of La$_{1-x}$Ca$_{x}$MnO$_{3}$}

\author{A.~Sani}
\affiliation{European Synchrotron Radiation Facility, BP 220,
F-38043 Grenoble, France}

\author{C.~Meneghini}
\email[Corresponding author: ]{meneghini@fis.uniroma3.it}
\affiliation{Dip. di Fisica Univ. di \emph{"Roma Tre"} Via della
vasca navale 84 I-00146 Roma Italy} \altaffiliation{INFM-OGG c/o
ESRF Grenoble, France}

\author{S.~Mobilio}
\affiliation{Dip. di Fisica Univ. di \emph{"Roma Tre"} Via della
vasca navale 84 I-00146 Roma Italy} \altaffiliation{Laboratori
Nazionali di Frascati dell'INFN, Via E. Fermi 40, I-00044 Frascati
Italy}

\author{Sugata Ray} \affiliation{Solid State and
Structural Chemistry Unit, Indian Institute of Science, Bangalore
560012, India}

\author{D.~D.~Sarma}
\affiliation{Solid State and Structural Chemistry Unit, Indian
Institute of Science, Bangalore 560012, India}

 \author{J.~A.~Alonso}
\affiliation{Instituto de Ciencia de Materiales, CSIC, Catoblanco,
Madrid, E-28049, Spain}

\date{\today}

\begin{abstract} The structure
of La$_{1-x}$Ca$_{x}$MnO$_{3}$ solid solutions (x=0, 0.25, 0.50,
0.67, 1) under high pressure (up to 40-45 GPa) has been
investigated by synchrotron X-ray powder diffraction (XRD) in
order to characterize their volume \emph{vs.} pressure (P-V)
equation of state. All the members of the solid solution, except
the extreme compounds: LaMnO$_3$ and CaMnO$_{3}$, present low
pressure orthorhombic P$nma$ phase evolving toward an high
pressure tetragonal (I$4/mcm$) symmetry. The details of this
transition are related to the  Ca doping: on increasing the Ca
content, the critical pressure of the transition decreases, but
the energetic cost of the transition increases. This study depicts
a peculiar evolution of structural distortions as a function of
pressure and concentration.

\end{abstract}
%      {62.50. +p}{High pressure and shock waves effects in solids and liquids}   \and
%      {61.50. Ks}{Crystallographic aspects of phase transformations; pressure effects} \and
%      {61.10. -i}{X-ray diffraction and scattering}

\maketitle
\section{Introduction}\label{intro}

Manganese oxides with perovskite structure form solid solutions of
the type: (\emph{Re}MnO$_3$)$_{1-x}$(\emph{Me}MnO$_3$)$_x$, in
which trivalent rare earth ions (\emph{Re} = La, Pr, Y) are
progressively substituted with divalent metal ions (\emph{Me} =
Ca, Ba, Sr). This substitution produces a mixed-valence state of
Mn$^{3+}$ and Mn$^{4+}$ ions giving rise to a phase diagram rich
in interesting transitions and phenomena such as metal to
insulator transition (MI), magnetoresistance (MR) and charge
ordering (CO) effects. Some of these oxides, in a range of
composition, are ferromagnetic metals at low temperatures and
become insulator above the Curie temperature (T$_c$). Crossing
T$_c$ is often associated with a large magnetoresistance effect
that stimulated the research interest since the early 1950's
\cite{Zener51}. In the last decade the discovery of huge
magnetoresistance in some of these compounds, named Colossal
Magnetoresistance (CMR), renewed intense attention on these
materials; besides their potential applications, these compounds
are intriguing from a fundamental point of view in the field of
strongly correlated systems.

The special features of manganese oxide perovskites originate from
a delicate balance among structural, magnetic and electronic
degrees of freedom whose main ingredients are: the double exchange
(DE) and super-exchange (SE) interactions between Mn ions, plus
 the coupling between charge carrier and lattice
distortions driven by the Jahn-Teller effect on Mn$^{3+}$ ions.
The concept of DE was introduced
\cite{Zener51,Anderson55,Gennes60} long ago to describe the
essential interactions among Mn ions giving rise to the MR effect:
the DE, in fact, is dependent on the electron hopping probability
\emph{t}, between Mn$^{3+}$ and Mn$^{4+}$ ions. The \emph{t} is
affected by the relative alignment of Mn$^{4+}$/Mn$^{3+}$ core
spin ($t_{2g}$) owing to the strong Hund coupling between $t_{2g}$
and $e_g$ electron spins. Thus, an external magnetic field which
aligns the $t_{2g}$ spins, promotes the $e_g$ charge hopping,
producing a negative MR effect. The relevance of charge to lattice
coupling (e-p), driven by the Jahn-Teller effect on Mn$^{3+}$
ions, has been recognized only recently\cite{Millis95,Millis96}:
it provides the localization energy to drive the metallic state to
an insulator below the room temperature and is essential to
explain the huge MR effect. Finally the SE coupling, that is
responsible for antiferromagnetic (AFM) interaction between the
$t_{2g}^3$ states at neighboring sites, is a fundamental
ingredient stabilizing charge ordered phases in the $x\geq 0.5$
compounds.

The La$_{1-x}$Ca$_{x}$MnO$_{3}$ series of solid solutions is the
prototype of mixed-valence perovskites, with similar ionic sizes
of Ca and La allowing one to obtain solid solution over the entire
concentration range. The phase diagram of
La$_{1-x}$Ca$_{x}$MnO$_{3}$, in terms of magnetic, electronic and
structural properties as a function of temperature and composition
appears complex \cite{Schiffer95}: around (and above) room
temperature (RT) all the compounds are paramagnetic insulators
(PM-I). Ca doped compounds, in the $\sim 0.2 \leq x < 0.5$
composition range, exhibit a paramagnetic-insulator (PM-I) to
ferromagnetic-metallic (FM-M) phase transition upon cooling. The
transition temperatures (T$_c$-T$_{MI}$) can be tuned by applying
a magnetic field and this gives rise to the so called CMR effect.
In the La doped compounds (that is in the $0.5 \leq x <0.85$
range) the prevalence of the AFM super-exchange coupling among Mn
ions results in low temperature charge ordered (CO) phases that
are insulating due to the reduced DE coupling\cite{Fernandez99}.
In a narrow region of composition around $x = 0.5$ the competition
between the FM double exchange and the AFM super-exchange coupling
makes the magnetic, electronic and structural phase diagram highly
intricate with the appearance of complex charge ordered
superstructures \cite{Ohsawa01}.

It is well-known that the atomic structure of these systems
influences the electronic and magnetic structures in diverse ways.
From the structural point of view LaMnO$_3$ (Mn$^{3+}$ ions) is
only made of Jahn-Teller (JT) distorted MnO$_6$ octahedra while,
in the CaMnO$_3$ (Mn$^{4+}$ ions), the MnO$_6$ octahedra are
undistorted. The doped compounds show a complex behaviour as a
function of temperature and doping: large JT effect is observed in
the high temperature PM-I phase, whose amplitude decreases with
increasing the Ca content
\cite{Booth98,Radaelli96,Louca99,Louca01}. A sudden drop of
MnO$_6$ JT distortions accompanies the insulator to metal
transition on cooling the sample ($0.2 \leq x < 0.5$) across
T$_c$-T$_{MI}$ \cite{Booth98,Lanzara98,Meneghini02,Subias02}. The
low temperature CO phases observed in La doped oxides
(x~$\geq$~0.5) are characterized by large cooperative JT effect
enhancing the MnO$_6$ coherent distortions \cite{Bardelli02}. Not
only the long range geometric structure, but also the local
structural aspects over the short and medium ranges have profound
influence on the physical properties of Mn perovskites, as they
determine the relative strengths of various interactions involved.
Thus, structural deformation, which modifies the distance between
Mn ions, the Mn-O-Mn bond angle and the cell volume, strongly
affects the physical properties of the compound such as the
electrical conductibility, the magnetic response, T$_c$, T$_{MI}$
and so on\cite{Hwang95b,Yoshizawa98}.

Relevant information on the physics of these materials can be
derived from studies under high
pressure~\cite{Hwang95,Nossov98,Neumeier95,Tokura96,Laukhin97,Senis98}
since squeezing the structure is a way to alter the scale of the
interactions involved. Studies as a function of hydrostatic
pressure, in the relatively low pressure region (up to 2-3 GPa)
reported the increase of charge delocalization
~\cite{Hwang95,Laukhin97} and suggested, in accordance with
chemical pressure results, a progressive metallization of the
system in the higher pressure region. However, recently,
complementary observations from Raman\cite{Congeduti01} and
IR\cite{Postorino01} spectroscopy, as well as electrical
conductibility and X-ray dif\-frac\-tion (XRD)\cite{MeneghiniHP}
on La$_{0.75}$Ca$_{0.25}$MnO$_3$ under higher pressures (up to
about 12 GPa) have reported the occurrence of a competing
mechanism which is activated raising the pressure above 6-7 GPa.
This mechanism appears to compete with the charge delocalization
preventing the metallization of the system raising the pressure at
least till 12 GPa. These recent findings strongly suggest the need
for a systematic investigation of the high pressure properties of
these compounds.

In this work we describe the structural phase diagram of
La$_{1-x}$Ca$_{x}$MnO$_{3}$ compounds as a function of pressure up
to about 40 GPa and composition ($0 \leq x \leq 1$), thereby
providing a complete description of this interesting class of
systems. The crystallographic structure of the samples has been
studied by synchrotron radiation XRD technique.

\section{Experimental and data analysis}\label{exp}

La$_{1-x}$Ca$_{x}$MnO$_{3}$ samples were prepared by conventional
solid-state reaction method using La$_2$O$_3$, CaCO$_3$ and
Mn$_3$O$_{4}$ precursors mixed in stoichiometric proportions. Soft
chemistry procedures have been employed to prepare the
stoichiometric LaMnO$_{3}$ and CaMnO$_{3}$ oxides. Stoichiometric
amounts of analytical grade La$_2$O$_3$, Ca(NO$_{3}$)$_{2}$ and
MnCO$_{3}$ were dissolved in citric acid. The citrate solutions
were slowly evaporated, leading to organic resins containing a
random distribution of the involved cations at an atomic level.
All the organic materials were eliminated in a treatment at
700$^{\circ}$C in air, for 12 hours. This treatment gave rise to
highly reactive precursor materials, amorphous to X-ray
diffraction (XRD). LaMnO$_3$ precursor was then treated at
1100$^{\circ}$C in a N$_{2}$ flow for 12 hours: inert-atmosphere
annealing was required to avoid the formation of oxidized
LaMnO$_{3+\delta}$ phases (rhombohedral in symmetry), thus
minimizing the unwanted presence of Mn$^{4+}$. CaMnO$_{3}$
precursor was annealed in air at 1100$^{\circ}$C for 12 h. All the
compounds appear single phase at standard X-ray diffraction
analysis.

High pressure X-ray Diffraction experiments were performed at the
high pressure beam line (ID9) of the European Synchrotron
Radiation Facility (ESRF) in Grenoble (France). Powder diffraction
patterns were collected using an angle dispersed set-up based on a
MAR345 imaging plate detector. The monochromatic beam  ($\lambda =
.41436$ \AA) was focussed to $30 \times 30~\mu m ^{2}$ spot using
a Pt-coated mirror (vertical focus) and an asymmetrically cut bent
Si (111) Laue monochromator (horizontal focus) \cite{zonton}.
Hydrostatic compression was achieved in a "Le Toullec" type
diamond anvil cell (DAC). The diameter of the diamond culet was
300 $\mu m$, smaller than in our previous experiment
(\cite{MeneghiniHP}) but it is mandatory in order to reach higher
pressures. A stainless steel gasket with a 125 $\mu m$ hole
diameter was used. Finely ground sample powder was placed in the
gasket hole using nitrogen as pressure transmitting media. The
diamonds, mounted with a thrust axis parallel to the incident
beam, allowed collecting a diffraction cone of about $25^\circ$ in
$2\theta$. The collection time was fixed to 5 s for each sample.
The DAC oscillates in between $\pm 3^\circ$. during acquisition in
order to improve the statistics over the sample orientations. The
pressure on the sample was determined before and after each
imaging, by monitoring the fluorescence line shift of a small ruby
pellet enclosed in the DAC with the sample.~\cite{Mao86} XRD
patterns were collected at room temperature in the pressure range
0-40~GPa with steps of about 1~GPa. 2D-XRD patterns were collected
on a MAR345 Imaging Plate (IP) with a pixel dimension of $100
\times 100$~$\mu m ^{2}$ placed about 300 mm from the sample. With
this set-up, the angular resolution calibrated on Si-NBS, was of
about 0.04$^\circ$ FWHM in 2$\theta$. The images were treated
using the FIT2D package~\cite{FIT2D} in order to have standard
intensity vs. $2\theta$ patterns, corrected for geometrical
effects.  Structural refinement of integrated profile patterns was
achieved through the Rietveld method as implemented in the GSAS
package.~\cite{GSAS} Pseudo-Voigt profile function proved to be
appropriate to fit the experimental pattern without asymmetry
correction. The background was modelled using a 12 term expansion
of the Chebychev polynomial function.

\begin{figure}
\begin{center}
\includegraphics[width=7cm,clip=]{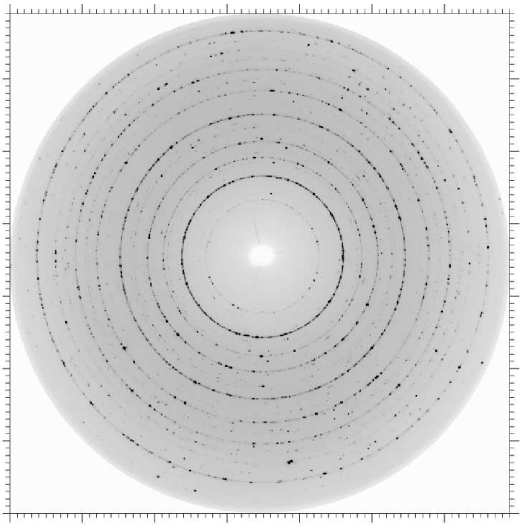}
\end{center}
\caption{\label{fig:01n}Typical 2D XRD pattern as collected using
the MAR345 detector.}
\end{figure}

In figure \ref{fig:01n}, a raw 2D XRD pattern image as collected
from a perovskite sample is presented. The very reduced sample
volume  probed by the XRD beam (being $\approx 10^{-9} cm^3$)
produces large orientational effect as shown by the inhomogeneous
intensity distribution along the Debye rings. It is clear that the
angular dispersed set-up and the 2D detector are mandatory for
such a kind of experiments since it represents the only way to
recover the preferred orientation effect and obtain a reliable
pattern intensity. In addition the MAR345 detector allows
collecting data of high quality thanks to the high count
statistics, to its linearity and to the rapid acquisition time
avoiding for pressure gradients during collection. It is important
to note that this detector allows reading in situ the 2D patterns,
this avoids incertitude in the experimental geometry calibration
(beam center, sample to detector distances, detector tilts) that,
on the contrary, may cause larger problems using ex-situ IP
read-out devices.

\begin{figure}
\begin{center}
\includegraphics[width=8cm]{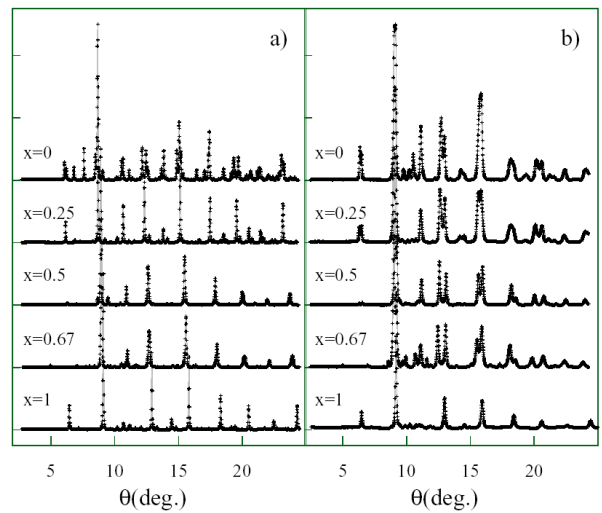}
\end{center}
\caption{\label{fig:02n} XRD patterns for the five samples
investigated at low pressure (P$\approx 3 GPa$, panel a) depict
the same P$nma$ space group symmetry. The high pressure phases
(P$\approx 30 GPa$, panel b) are function of the sample
composition: x=0 sample present Imma orthorhombic phase, x=0.25,
0.5 and 0.67 have tetragonal I$4/mcm$ high pressure phase, x=1 has
orthorhombic P$nma$ phase in the whole pressure range
investigated. }
\end{figure}

Before we proceed to the data presentation and to the discussion
of the results it is worth while to notice that though the
absolute accuracy of XRD pattern analysis can be affected by the
calibration procedure (beam wavelength, sample to detector
distance and so on), the relative modifications of XRD patterns
induced by the pressure is reliable being largely independent of
such factors. Thus, structural evolution under pressure and, more
in general, relative changes in structural features, can be
followed with very high accuracy with this set up, arising from
the exceptional statistics and the integration over the whole
Debye rings allowing also weak pattern modifications to be
recognized.

 The equation of state has been
quantitatively analyzed by fitting the V \emph{vs.} P curves with
the Birch-Murnaghan (B-M) equation of state \cite{Birch47}. The
B-M equation is based on the assumption that the strain energy can
be expressed as a Taylor series in the finite strain \emph{f}.
There are several alternative definitions for \emph{f}
corresponding to different equation of state, each one leading to
a different relationship between P and V. The B-M equation is
based on the so called Eulerian definition of \emph{f} that is:
\begin{eqnarray}\label{fe}
f_E=\frac{1}{2}\left[(V_0/V)^{2/3}-1\right]
\end{eqnarray}
and the expansion to the fourth order in the strain yields:
\begin{eqnarray}\label{bm}
 P = 3 B_0 f_E(1+2f_E)^{5/2}
 \left[1+\frac{3}{2}(B'-4)f_E + \right. \;\;\;\;\;\;\;\;\; \nonumber \\
 \left.\;\;\;\;\;\;\;\;\;\;\frac{3}{2}
 \left(B_0 B''+(B'-4)(B'-3)+\frac{35}{9}
       \right) f_E^2
       \right]
\end{eqnarray}

The bulk modulus at the zero pressure ($B_0=-V \partial P/\partial
V$) and its pressure derivatives ($B'= \partial B/\partial P$ and
$B''= \partial^2 B/\partial P^2$) characterize the stiffness of
the solid. If the equation is truncated at the second order in the
energy, then the coefficient of $f_E$ must be identical to zero,
thus $B'=4$. The truncation to the third order, that implies the
coefficient of $f_E^2$ equals zero, gives a three parameters
equation: $V_0$, $B_0$ and $B'$,  and implies $B_0''= -1/B_0 [
(3-B')(4-B')+35/9 ]$. Within this formalism the normalized
pressure is defined as: $F_E=P[3f_E(1+2f_E)^{5/2}]^{-1}$ so that
the plot $F_E$ \emph{vs.} $f_E$ gives an intuitive way to
determine the order of the B-M equation: a constant $F_E(f_E)$
implies a second order B-M equation of state, a linear trend
requires a third order B-M equation and quadratic trend justifies
the fitting with a fourth order equation of state.

\begin{table}
\caption{\label{t:lamno}Refined structural parameters from
synchrotron radiation X-ray diffraction for LaMnO$_{3}$ at
selected pressures in the $Pnma$ and $Imma$ space groups. Numbers
in parentheses are statistical errors at the last significant
digit. The atomic positions in the $Pnma$ are: Mn (0.,0.,0.5); La
and O(1) (\emph{x}, 0.25, \emph{z}); O(2)(\emph{x}, \emph{y},
\emph{z}). In the $Imma$ space group the atomic positions are: Mn
(0., 0., 0.5); La and O(1) (0., 0.25, \emph{z}) and O(2) (0.25,
\emph{y}, 0.75).}
\begin{tabular}{lllll}
\hline\noalign{\smallskip}
                  &      &2.1~GPa   & 16.3 GPa  & 28.2 GPa \\
                  &      &(P$nma$)  & (I$mma$)  &  (I$mma$)\\
\noalign{\smallskip}\hline\noalign{\smallskip}
\em{a} (\AA)      &      & 5.6349(4)& 5.2703(8) & 5.320(2)\\
\em{b} (\AA)      &      &7.6912(5) & 7.522(1)  & 7.391(2)\\
\em{c} (\AA)      &      & 5.5233(4)& 5.47323(5)& 5.362(2)\\
\em{V} (\AA$^{3}$)&      & 329.28(2)& 221.10    & 218.83\\
La                &\em{x}& 0.0430(5)&           & \\
                  &\em{z}& 0.9999(1)&-0.001(2)  & 0.017(1)\\
O(1)              &\em{x}& 0.497(6) &           & \\
                  &\em{z}& 0.088(6) & 0.083(3)  & -0.071(8)\\
O(2)              &\em{x}& 0.293(5) &           & \\
                  &\em{y}& 0.034(3) & 0.054(2)  & -0.035(4)\\
                  &\em{z}& 0.714(7) &           & \\
R$_{WP}\;\%$      &      &  2.53    & 2.30      & 2.90\\
$\chi^{2}$        &      &    0.7   &   0.6     & 0.6\\
\noalign{\smallskip}\hline
\end{tabular}
% Or use
%\vspace*{5cm}  % with the correct table height
\end{table}

\begin{figure}[hbt]
\begin{center}
\includegraphics[width=5cm]{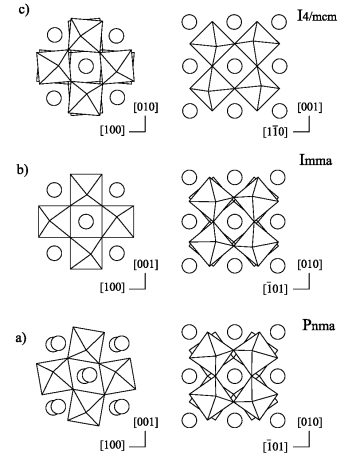}
\end{center}
\caption{\label{fig:03n} The main structural features found in the
high pressure phase diagram of La$_{1-x}$Ca$_x$MnO$_3$ samples:
the orthorhombic P$nma$  structure (a) is characterized by three
independent Mn-O distances and three tilt modes. The I$mma$
structure (b) has two independent Mn-O distances and two two tilt
modes. The tetragonal I$4/mcm$ structure (c) has two independent
Mn-O distances and one tilt mode.}
\end{figure}

\section{Results}\label{res}

The figure \ref{fig:02n} reports examples of low and high pressure
XRD patterns for the investigated samples as a function of
composition. Despite the Ca doping, $0\leq x \leq 1$, all the
members of the family, at low pressure, represent an orthorhombic
structure described by the P$nma$ space group. Such a structure is
derived from the ideal cubic perovskite structure in which the
MnO$_{6}$ octahedra, are located on the cube corners, while the
La/Ca ions occupy the cube centers. However, since the La/Ca ions
are too small to fill the free space inside the cube (size
mismatch), the space filling is achieved mainly by tilting the
MnO$_{6}$ octahedra with respect to the crystallographic planes,
thus deforming the cube and doubling the unit cell along the b
axis (figure \ref{fig:03n}a). Within the P$nma$ space group
description, the lattice parameters (a, b, c) are related to the
edge of the ideal perovskite cubic cell $a_{p}$, by $a = c \sim
\sqrt 2 a_{p}$ and $b=2a_{p}$. The two inequivalent oxygen sites
in the $Pnma$ structure also allow taking into account the JT
distortion of MnO$_6$ octahedra observed for $x<1$. High pressure
structures are found to depend on the Ca doping: the structure of
LaMnO$_3$ remains orthorhombic, but the space group changes from
P$nma$ to I$mma$ characterized by alternate rotation of the
octahedra around the [101] direction (figure \ref{fig:03n}b). The
high pressure phase in doped compounds ($0<x<1$) is tetragonal
(I$4/mcm$), characterized by alternate rotation of the octahedra
around the [001] axis (figure \ref{fig:03n}c). The structure of
CaMnO$_3$ remains orthorhombic in the P$nma$ space group in the
whole pressure range investigated.

The structural evolution of the samples as a function of pressure
and composition is detailed in the following sections.

\subsection{LaMnO$_{3}$}\label{1}

\begin{figure}[hbt]
\begin{center}
\includegraphics[width=5.5cm]{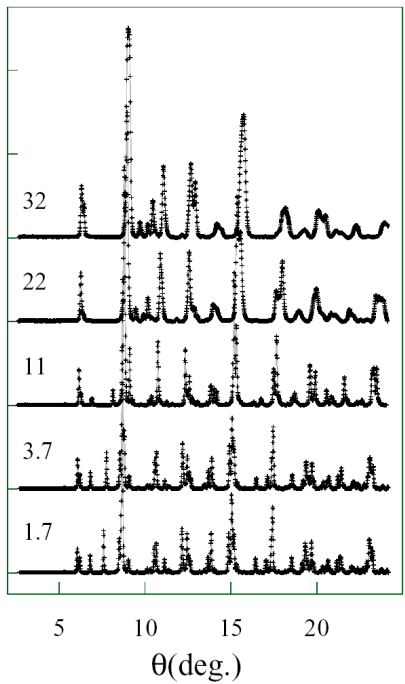}
\end{center}
 \caption{\label{fig:04n} Selected XRD patterns
of LaMnO$_{3}$ as a function of applied pressure collected at RT.
Pressure are reported in GPa.}
      % Give a unique label
\end{figure}

\begin{figure}[htb]
\begin{center}
\includegraphics[width=7cm,clip=]{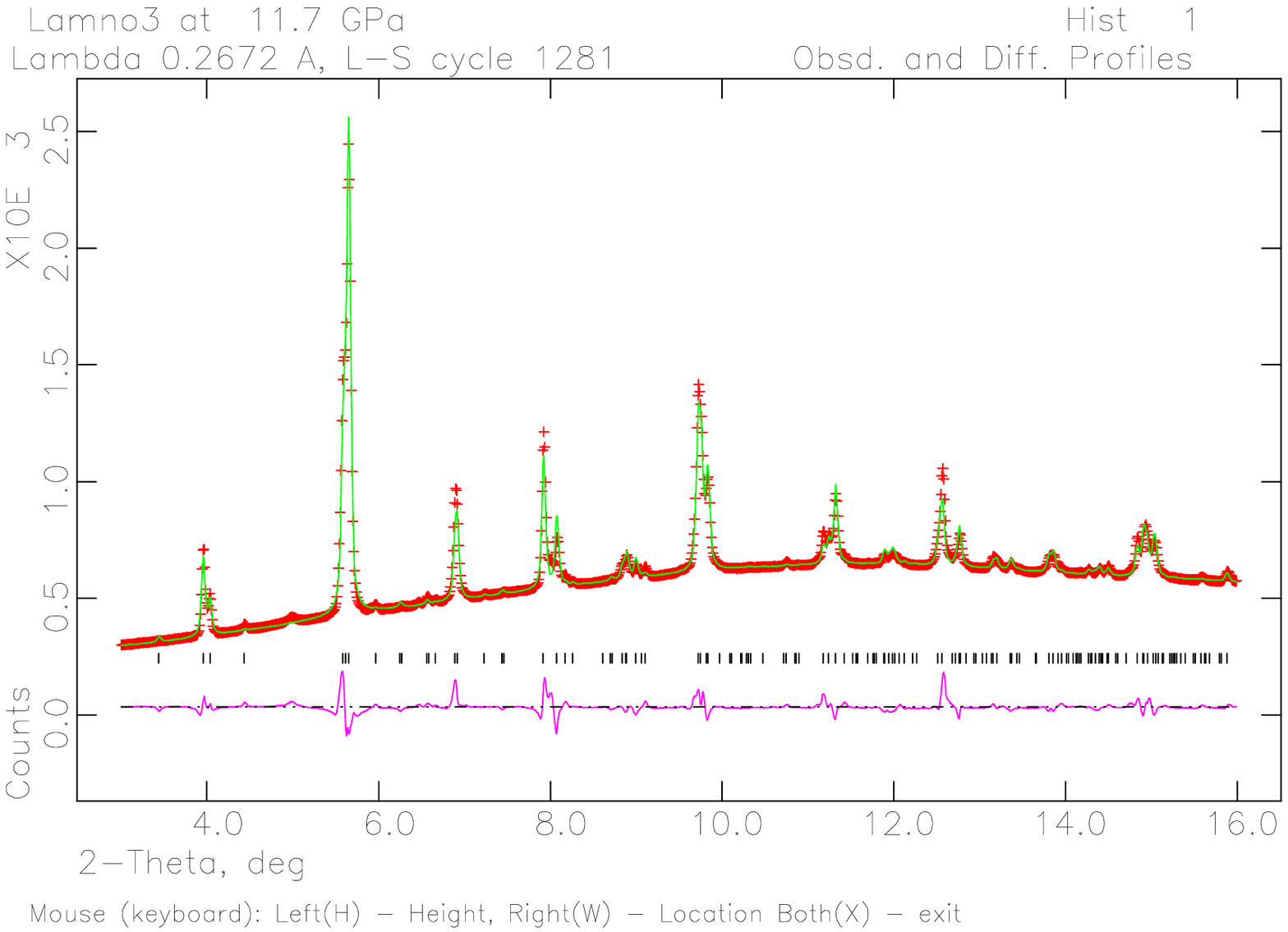}
\end{center}
\caption{Example of XRD Rietveld profile refinement on LaMnO$_{3}$
pattern collected at 11.7 GPa. Experimental data (crosses) best
fit profile (line) and residual (shifted for clarity) are shown.
Markers represent the expected positions for the diffraction
lines. }
\label{fig:05n}       % Give a unique label
\end{figure}

Typical diffraction patterns of LaMnO$_{3}$ in the pressure range
1-32 GPa are shown in figure~\ref{fig:04n} and an example of the
profile fitting (P=11.7 GPa) is reported in figure~\ref{fig:05n}.
The good data quality allows a satisfactory refinement up to P =
32 GPa yielding reliable lattice parameters and atomic positions.

The figure \ref{fig:06n}(0.0-a) reports the evolution of
LaMnO$_{3}$ lattice edges as a function of pressure. The $b$-axis
values reported in the figure are normalized by a factor
$1/\sqrt{2}$, corresponding to a pseudo-cubic representation of
the lattice. The compressibility is different along the three axes
and evolves differently as a function of the applied pressure as
shown by the data reported in figure \ref{fig:06n}(0.0-a): the
LaMnO$_{3}$ $a$-axis represents the soft compressibility direction
up to about 16~GPa, thus the $a$-axis compressibility decreases.
The $b$-axis compressibility changes (slightly decreases) above
$\approx 20$ GPa.  Along the $c$-axis, the compressibility
decreases smoothly till about 14 GPa and then it abruptly
increases above $\approx 14$ GPa. This trend is in qualitative
agreement with that reported recently\cite{Ingo01}. The structure
of LaMnO$_{3}$ remains orthorhombic up to 35 GPa.

However the changes of compressibility along the three axes in the
region 10-20~GPa, are indicative of a phase transition and the
progressive decrease of intensity of the (1 1 1) reflection (see
figure~\ref{fig:04n}) is suggestive of a $I$ space group, since
the reflections with \emph{h+k+l} odd are forbidden by this
symmetry. This qualitative hypothesis is further confirmed by the
impossibility to achieve satisfactory structural refinement
convergence within the P$nma$ space group above $\approx$~15~GPa.
The refinements of XRD profiles demonstrate a change of space
group from $Pnma$ to $Imma$ above $\approx 15$~GPa, as evidenced
by the volume compressibility change visible in
figure~\ref{fig:06n}(0.0-b). The $Imma$ space group is derived
from the ideal cubic perovskite structure by a single rotation of
the octahedra around the [1 0 1] cubic direction
(figure~\ref{fig:03n}). This phase has been already observed in
manganese oxide perovskites containing cations with a large ionic
radius, such as La$_{0.70}$Ba$_{0.30}$MnO$_{3}$
\cite{radaelli96b}, as a result of chemical substitution effect.
However, P$nma$-I$mma$ transition in the present case of physical
pressure does not involve any intermediate trigonal phase that is
observed in studies performed as a function of the chemical
pressure\cite{radaelli96b}. The smoothness in the evolution of the
cell parameters and cell volume as a function of applied pressure
(figure \ref{fig:06n}(0.0-a,b)) demonstrate the continuous (second
order) nature of this phase transition. The V \emph{vs.} P curve
is well described by a third order Birch-Murnaghan equation of
state \cite{Birch86}, yielding the values reported in table
\ref{tab:1} for the bulk moduli, $B_o$ and $B'$. These values are
in agreement with those reported in literature\cite{Ingo01}. The
values of $B'$ (that is 9.5 in the P$nma$ phase and 12 in the
I$mma$ phase) are large in comparison to those typical of crystals
with nearly isotropic compression that ranges in between 4-6.
These large values found here must be related to the anisotropic
compressibility along the three axes, as already suggested in
ref.\cite{Ingo01}. The bulk modulus and its derivative increase
through the P$nma$ to I$mma$ transition pointing out the
increasing stiffness of the structure. The P$nma$ to I$mma$
transition increases the stiffness of the structure with the
doubling of the bulk modulus, from about 100 GPa to 200 GPa.

We can clearly notice from fig. 6a that $a$ and $c$ axes lengths
approach each other raising the pressure toward the maximum value
(32 GPa). This trend, in analogy with the results obtained on the
other samples (see below), may suggest an impending orthorhombic
to tetragonal phase transition which might be reached at somewhat
higher pressure.

The good quality of the patterns allows reliable refinement of the
atomic positions from which the Mn-O bond distances are
calculated, as reported in figure~\ref{fig:06n}(0.0-c). In the
P$nma$ phase, the three Mn-O distances remain distinguishable,
above the statistical uncertainty, till about 14 GPa. This
finding, in agreement with recent neutron diffraction
results~\cite{Pinsard01}, demonstrates the stability of JT effect
even in the high pressure region. The P$nma$ to I$mma$ transition
provokes the collapse of the three Mn-O distances into one single
value, representing removal of the coherent JT distortion.

The orthorhombic strains in the $ac$ plane,
$\rm{Os}_{\parallel}=2(a-c)/(a+c)$, and along the $b$ axis with
respect to the $ac$ plane,
$\rm{Os}_{\perp}=2(b/\sqrt(2)-(a+c)/2)/(b/\sqrt(2)+(a+c)/2)$, are
reported in figure~\ref{fig:06n}(0.0-d). They represent the
deviation of the structure from an ideal cubic cell.
$\rm{Os}_{\parallel}$ is quite large ($\sim$2\%) at low pressures
decreases to zero ($a=b$) around 10 GPa. Thus it increases in
absolute value reaching the -2\% in between 15 and 25 GPa. The
decreasing trend (in absolute value) of $\rm{Os}_{\parallel}$
raising the pressure above 25 GPa is one other finding suggesting
an incoming tetragonal transition to be reached at higher pressure
(see the panel 0.25-d in figure~\ref{fig:06n}). The effect of
pressure on the $\rm{Os}_{\perp}$ is weaker: in the P$nma$ phase
it decreases (in absolute value) from -2.8\% to about -2\%, and
remains quite unchanged in the high pressure I$mma$ phase.

\begin{table}[htb]
\caption{\label{t:Ca25}Refined structural parameters from
refinement of   La$_{0.75}$Ca$_{0.25}$MnO$_{3}$ XRD patterns at
selected pressures in the $Pnma$, $Imma$ and $I_{4}/mcm$ space
groups. Numbers in parentheses are statistical errors at the last
significant digit. Example of structural refinement within the
$Imma$ space group ($P\sim 20 GPa$) is reported in the third
column. The atomic positions in the P$nma$ are: Mn (0., 0., 0.5);
La and O(1): (\emph{x}, 0.25, \emph{z}); O(2)
(\emph{x},\emph{y},\emph{z}). In the I$mma$ space group are: Mn
(0., 0., 0.5); La and O(1) (0., 0.25, \emph{z}) and O(2) (0.25,
\emph{y}, 0.75). In the $I_{4}/mcm$ space group the atomic
positions are: Mn (0., 0., 0.); La/Ca and O(1)
 (0., 0., 0.25) and O(2) (\emph{x}, \emph{x}+0.5, 0.).}
\begin{tabular}{llllll}
\hline\noalign{\smallskip}
                  &      &6.1~GPa    &  20.2 GPa &  27.1  GPa& 38.2 GPa     \\
                  &      &($Pnma$)   &  ($Imma$) &  ($I_{4}/mcm$)& ($I_{4}/mcm$)\\
\noalign{\smallskip}\hline\noalign{\smallskip}
\em{a} (\AA)      &      & 5.4133(3) & 5.3231(3) & 5.2846(6)& 5.2050(2)\\
\em{b} (\AA)      &      & 7.6500(5) & 7.4867(1) & 7.587(2) & 7.561(2)\\
\em{c} (\AA)      &      & 5.4478(3) & 5.3813(1) &          & \\
\em{V} (\AA$^{3}$)&      & 225.61(1) & 214.45    & 211.9    & 204.84\\
La/Ca             &\em{x}& 0.07(1)   &           &          & \\
                  &\em{z}&-0.001(1)  &-0.058(6)  &          & \\
O(1)              &\em{x}& 0.49(2)   &           &          & \\
                  &\em{z}& 0.113(6)  & 0.059(6)  &          &          \\
O(2)              &\em{x}& 0.28(1)   &           & 0.31(1)  & 0.32(1)   \\
                  &\em{y}& 0.033(4)  &-0.016(4)  &          & \\
                  &\em{z}& 0.718(8)  &           &          &  \\
R$_{WP}\;\%$      &      & 3.80      & 3.07      & 2.93     &  2.63\\
$\chi^{2}$        &      &    0.7    & 0.4       & 1.2      & 0.6\\
\noalign{\smallskip}\hline
\end{tabular}
\end{table}

\begin{table}[htb]
\caption{\label{tab:1} Bulk moduli at the zero pressure (B$_o$)
and their pressure derivatives (B' and B") derived by fitting the
V(P) data with the B-M equation of state.}
\begin{tabular}{lllll}
\hline\noalign{\smallskip}\multicolumn{2}{c}{La$_{1-x}$Ca$_{x}$MnO$_{3}$ }& $B_{0}$  & $B'$ & $B''$       \\
                            &  phase & (GPa)    & $B'$ & (GPa$^{-1}$)\\
\noalign{\smallskip}\hline\noalign{\smallskip}
x=0                         & P$nma$   & 104(2) & 9.5(2) & \\
                            & I$mma$   & 200(5) & 12.0(6)& \\
\noalign{\smallskip}\noalign{\smallskip}
x=0.25                      & P$nma$   & 136(4) & 10.2(3)& \\
                            & P$nma$*  & 176(6) & 3.3(1) & 0.9(3)\\
                            & I$4/mcm$ & 310(30)& 7(3)   & \\
\noalign{\smallskip} \noalign{\smallskip}
x=0.5                       & P$nma$   & 172(5) & 5.7(6) & \\
                            & I$4/mcm$ & 253(5) & 6.4(8) & \\
\noalign{\smallskip}\noalign{\smallskip}
x=0.67                      & P$nma$   & 210(7) & 2(1)   & \\
                            & I$4/mcm$ & 186(7) & 4(1)   & \\
\noalign{\smallskip}\noalign{\smallskip}
x=1                         & P$nma$*  & 195(3) & 2.9(2) & 0.2(1)\\
\noalign{\smallskip}\hline
\end{tabular}
\end{table}

\subsection{La$_{0.75}$Ca$_{0.25}$MnO$_{3}$}

\begin{figure*}[h]
\begin{center}
\includegraphics[width=4.2cm,clip=]{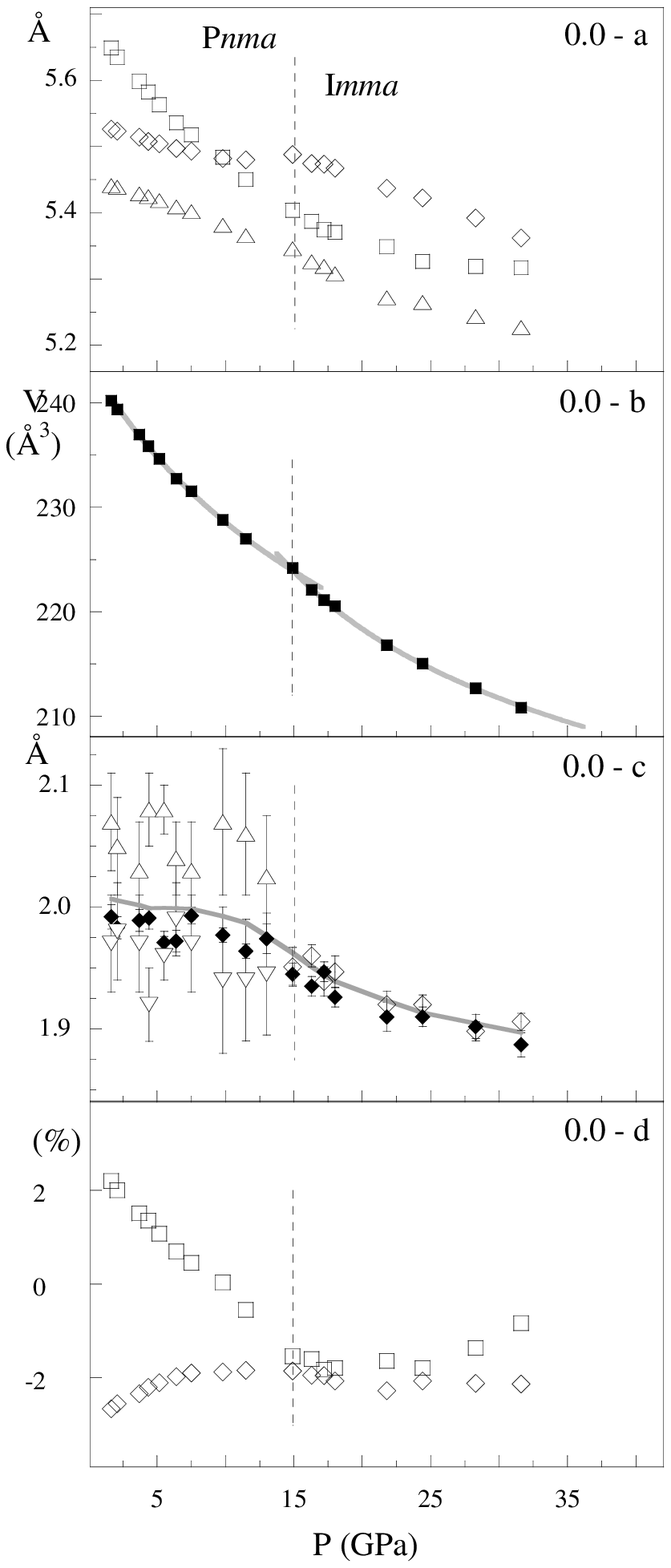}
\includegraphics[width=4.2cm,clip=]{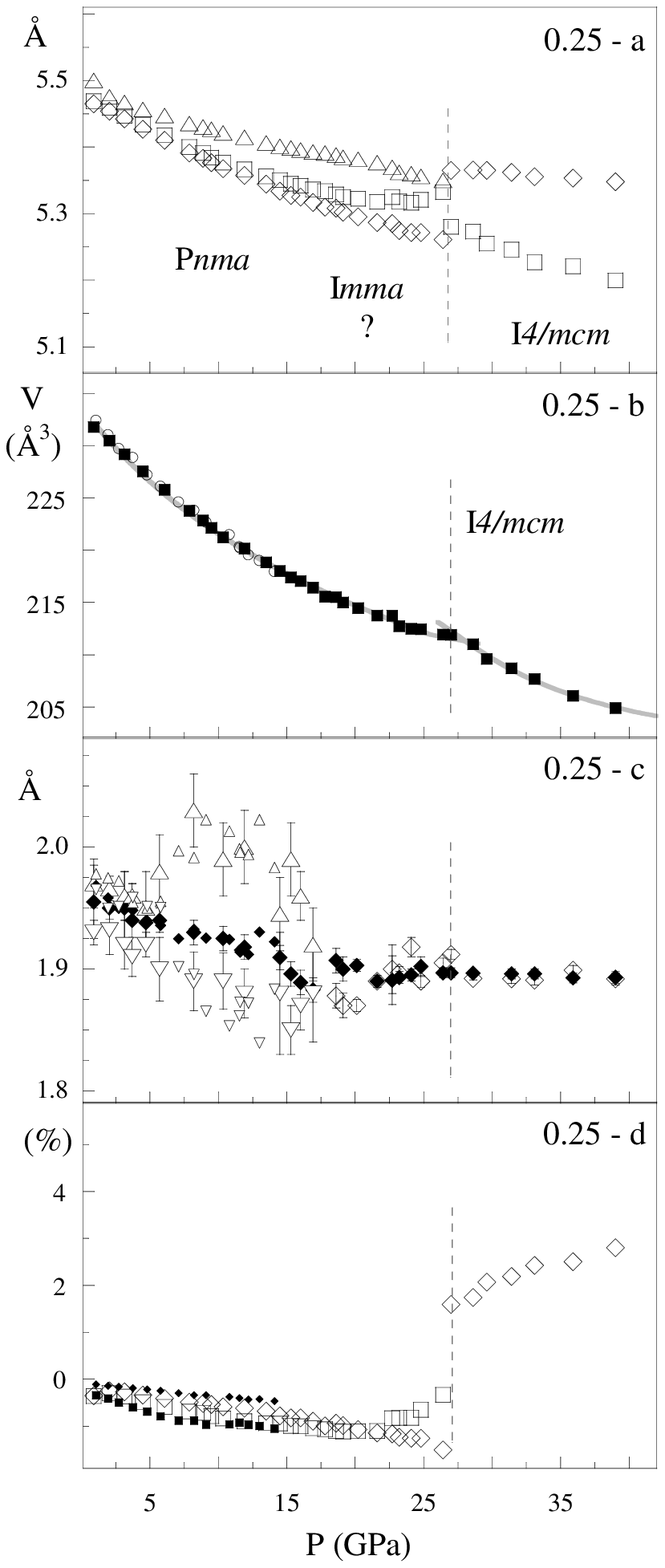}
\includegraphics[width=4.2cm,clip=]{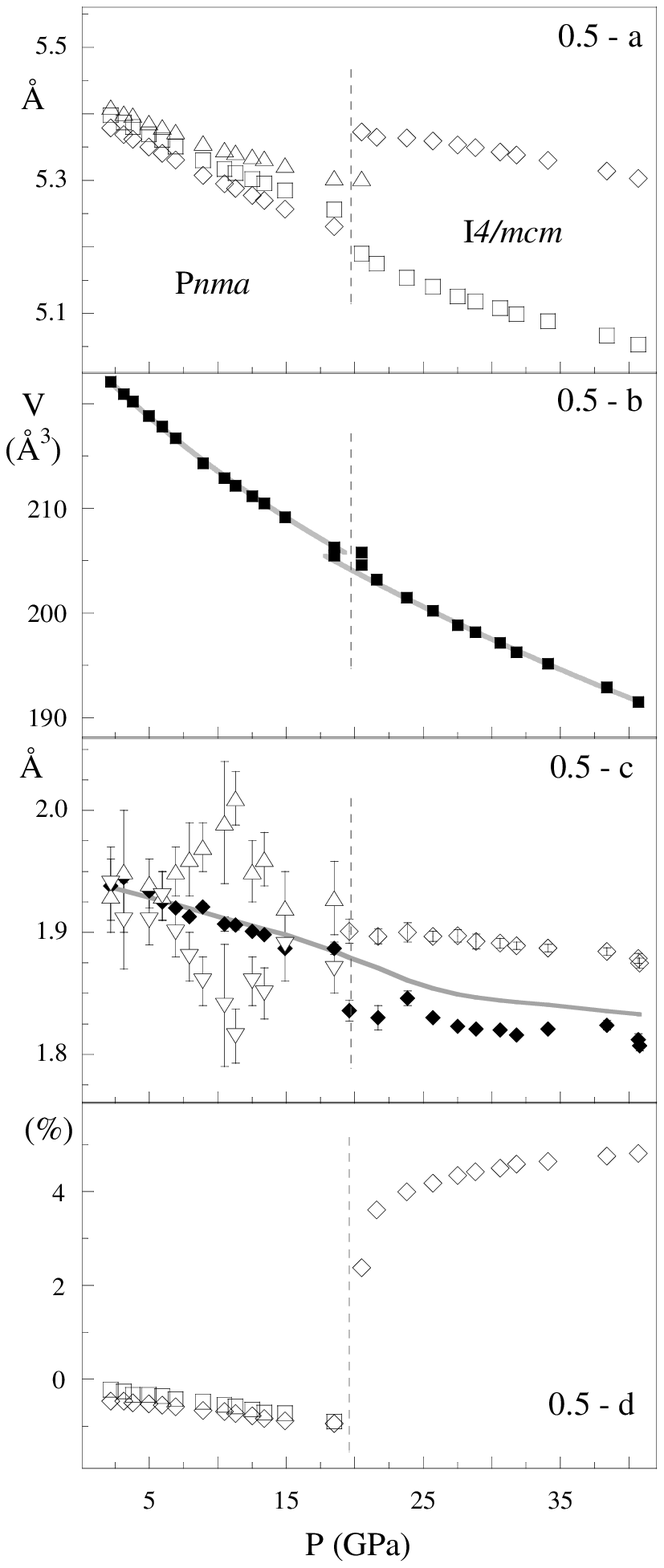}
\includegraphics[width=4.2cm,clip=]{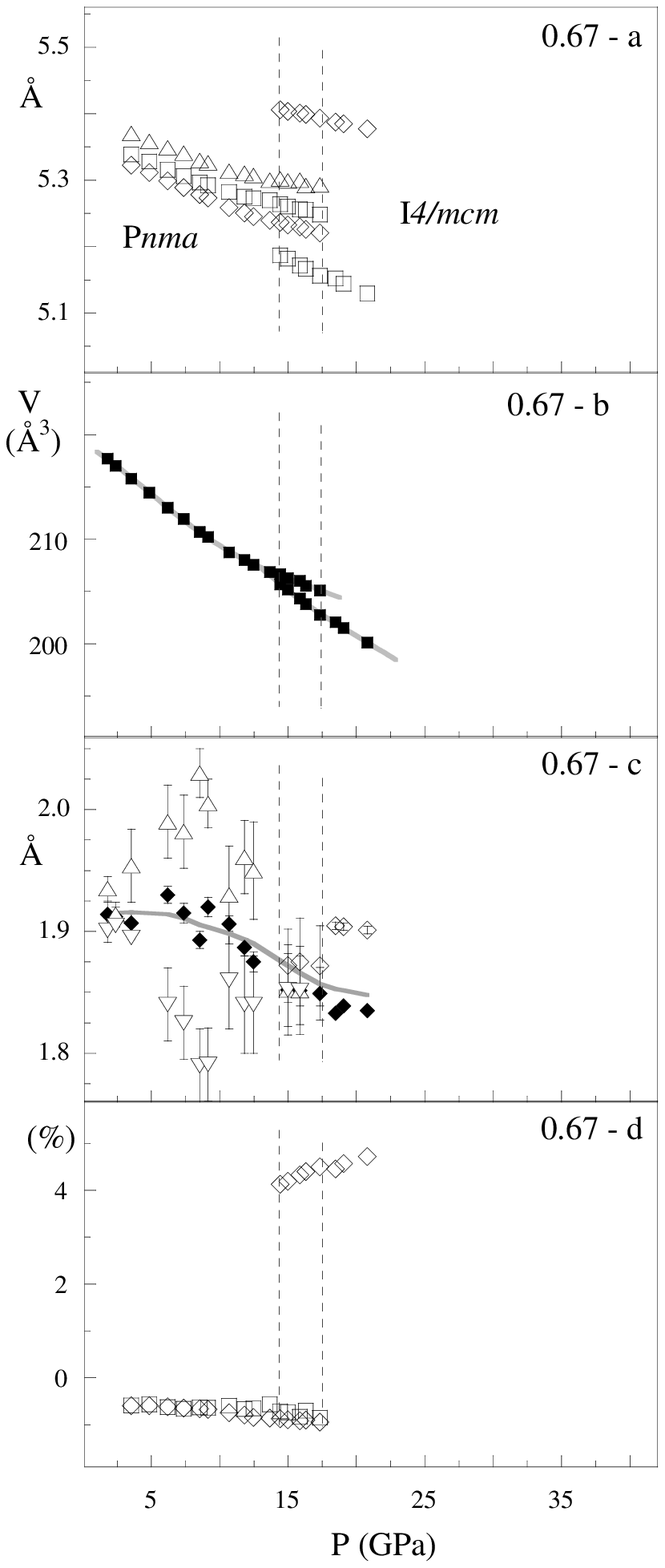}
\end{center}
\caption{The principal structural parameters for x=0 (0.0 plots),
x=0.25 (0.25 plots), x=0.5 (0.5 plots) and x=0.67 (0.67 plots)
samples as a function of the applied pressure. Panels \emph{a}
show the lattice parameters ($a$=squares, $b/\sqrt{2}$=diamonds
and $c$=triangles). The volume \emph{vs.} pressure data are
reported in panels \emph{b} (squares black), the continuous gray
lines refer to the fit of the Birch-Murnaghan equation of state.
Mn-O distances are shown in panels \emph{c}: three different Mn-O
distances are found in the P$nma$ phase and only two in the
tetragonal phase. The grey line is a smooth fit through the
average Mn-O distance points. Panels \emph{d} report the
orthorhombic (tetragonal) strain parallel (Os$_{\parallel}$) to
the ac plane (squares) and along $b$ with respect to the $ab$
plane (Os$_{\perp}$) (diamonds).  For sake of comparison the
previous results for x=0.25 sample are also reported in panels
0.25-b (open circles), 0.25-c (small symbols without error bars)
and 0.25-d (small diamonds).} \label{fig:06n}
\end{figure*}

The high pressure structural behavior of
La$_{0.75}$Ca$_{0.25}$MnO$_{3}$ was investigated by some of us in
a recent work\cite{MeneghiniHP} extending the pressure range up to
about 12 GPa. In this experiment we have been able to extend the
pressure range up to about 40 GPa by using a smaller diamond culet
in the DAC cell. Unfortunately, the quality of the data collected
in this sample is slightly poorer than in the previous experiment.
This is partially due to the smaller sample volume that reduces
the statistics in XRD patterns. For this reason we have a slightly
larger uncertainty refining the the lattice parameters and, in
particular, refining the Mn and O atomic positions. Notice,
however, that the new values are in agreement with those
previously reported in the overlapping range
(figure~\ref{fig:06n}(0.25)).

The lattice parameters reported in figure~\ref{fig:06n}(0.25-a)
depict a decreasing trend as a function of the applied pressure
till about 20 GPa. The easy compressibility axis above 10 GPa is
the $b-$axis. The evolution of the $b$-axis as a function of the
applied pressure is quite smooth till about 27 GPa. On the
contrary, we notice a smooth change of slope in the $a$ and
$c$-axes \emph{vs.} P curves above $\sim 10$ GPa. Changes in $a$-
and $c$-axes compressibility become more accentuated on raising
the pressure above $\sim 20$ GPa, causing the convergence of $a$
and $c$ lattice parameters in the region preceding to the
orthorhombic (P$nma$) to tetragonal (I$4/mcm$) phase transition
occurring around 27 GPa. We must notice that the intensity of the
(1 1 1) reflection in the XRD patterns becomes progressively
weaker on raising the pressure as we already observed in
LaMnO$_{3}$. This trend and the change of compressibility along
the $a$- and $c$-axes are compatible with a sluggish P$nma$ to
I$mma$ phase transition starting around 17-20~GPa. This hypothesis
is in agreement with the reported P$nma$ to I$4/mcm$ through
I$mma$ phase transition as a function of average cation
size\cite{Woodward98}. We must notice, in addition, that we found
a progressive increase of the $\chi^2$ factor refining our XRD
patterns above 10-13~GPa. This suggests that the progressive
appearance of other crystallographic phases, together with the
P$nma$ one, may even be starting at lower pressures. The
hypothesis of parasitic phases above $\sim 10$ GPa is also in
agreement with the quadratic trend observed in the plot of the
normalized pressure, $F_E$, as a function of the Eulerian strain
$f_E$ (see below).

The V(P) curve reported in figure \ref{fig:06n}(0.25-b) depicts an
evident change of slope around 27 GPa, in correspondence of the
orthorhombic to tetragonal phase transition. The fitting with a
third order B-M  equation of state~\cite{Birch86} in the P less
than as well as above 27 GPa allows extracting the values of the
bulk moduli in the P$nma$ and I$4/mcm$ phases, respectively. We
found $B_{0}$=136 GPa and $B'\sim 10$ in the orthorhombic phase,
and $B_{0}\approx 300$ GPa and $B'\sim 7$ in the high pressure
tetragonal phase (table \ref{tab:1}). The values of B$_{0}$ and
$B'$ are respectively smaller and larger in the P$nma$ phase than
those previously reported~\cite{MeneghiniHP}, namely $B_{0}$=178
GPa and $B'\sim 4$. The $B'$ value is anomalously high, similar to
that previously observed in the LaMnO$_{3}$ sample, even if, in
this sample the compressibility anisotropy along the three axes is
less evident. The figure \ref{fig:08n} shows the plot of the
normalized pressure, $F_E$, as a function of the Eulerian strain
$f_E$ for La$_{0.75}$Ca$_{0.25}$MnO$_{3}$ sample. In this plot, a
non linear trend in the $F_E(f_E)$ curve is evident, suggesting
the fitting with a fourth order B-M equation. In this case the
best fit gives the following values for the bulk modulus and its
derivatives: $B_{0}=176$ GPa , $B'=3.3$ and $B''=0.9$ GPa$^{-1}$.
Notice that in this case the values found for $B_0$ and $B'$ are
in agreement with our previous findings obtained over a restricted
pressure range, and the $B'$ is consistent with the more isotropic
compressibility of the structure. The quadratic trend found in the
$F_E$ \emph{vs.} $f_E$ curve could suggest the growing of
parasitic phases already around 10 GPa.

Around 25 GPa the orthorhombic structure becomes unstable and the
system evolves towards a tetragonal structure (I$4/mcm$). This
phase transition is characterized, below the critical pressure, by
a sort of plateau of the V(P) curve. The smoothness of the phase
transition is characterized by a progressive decrease of $c$,
which ``fuses'' with $a$. One may wonder why $b$ increases with
pressure, since the compression has commonly a squeezing effect on
the lattice parameters. This anomalous effect must stem from a
change in the tilt of the octahedra in the new phase. In fact the
effect of pressure on the structure is twofold: on one hand, the
pressure acts on the size of the Mn octahedra taking it closer to
a more symmetric structure characterized by one Mn-O distance. On
the other hand, the pressure induced phase transition modifies the
tilting of the octahedra (figure~\ref{fig:03n}). The P$nma$ phase
is characterized by a three-tilt system of the MnO$_6$
octhaedra~\cite{Glazer72} while in the tetragonal structure the
octahedra are aligned along the $b$ axis (Mn-O-Mn bond angle along
$b$ is 180$^\circ$) thus the cell must expand in this direction in
order to accommodate the two Mn-O bonds along $b$.

Even if the quality of the data makes the refinement of atomic
positions more uncertain with respect to the previous results, we
notice that the evolution of Mn-O bond lengths matches well our
previous results: at low pressures (P$<$5-6 GPa) the Mn-O bond
lengths appear indistinguishable within the statistical
uncertainty. Above 5-6 GPa three Mn-O distances are recognizable,
pointing out a larger coherent JT effect. The Mn-O bond lengths
collapse raising the pressure above $\sim$ 15 GPa, as the
structure evolves from the orthorhombic to the tetragonal phase.
The Mn-O bonds remain indistinguishable in the high pressure
orthorhombic region and in the tetragonal phase demonstrating the
absence of any coherent JT effect.

The orthorhombic strains (figure~\ref{fig:06n}(0.25-d)) are
sensibly lower than in the LaMnO$_3$ sample. Consistently with our
previous results both $\rm{Os}_{\perp}$ and $\rm{Os}_{\parallel}$
decreases (increases in absolute value) as a function of the
applied pressure. Raising the pressure above 20 GPa causes
$\rm{Os}_{\parallel}$ to approach zero before the transition to
the tetragonal phase. In the I$4/mcm$ phase $a=b$ causes
$\rm{Os}_{\parallel}=0$. The $\rm{Os}_{\perp}$ increases (in
absolute value) as a function of pressure in the whole range
investigated. The change of symmetry causes the change of sign
$\rm{Os}_{\perp}$ but its modulus evolves smoothly across the
phase transition.

\begin{table}[h]
\caption{\label{t:Ca50} Refined structural parameters from
refinement of  La$_{0.5}$Ca$_{0.5}$MnO$_{3}$ XRD patterns at
selected pressures in the $Pnma$ and $I_{4}/mcm$ space groups.
Numbers in parentheses are statistical errors at the last
significant digit. The atomic positions in the $Pnma$ are: Mn (0.,
0., 0.5); La and O(1): (\emph{x}, 0.25, \emph{z}); O(2)
(\emph{x},\emph{y},\emph{z}). In the $I_{4}/mcm$ space group the
atomic positions are: Mn (0., 0., 0.); La/Ca and O(1)
 (0., 0., 0.25) and O(2) (\emph{x}, \emph{x}+0.5, 0.).}
\begin{tabular}{lllll}
\hline\noalign{\smallskip}
                  &       & 5.9~GPa & 20.1 GPa     & 40.7 GPa     \\
                  &       & ($Pnma$)& ($I_{4}/mcm$)& ($I_{4}/mcm$)\\
\noalign{\smallskip}\hline\noalign{\smallskip}
\em{a} (\AA)      &      & 5.3605(2) & 5.1536(3) & 5.0533(5)\\
\em{b} (\AA)      &      & 7.5532(2) & 7.5853(7) & 7.499(1)\\
\em{c} (\AA)      &      & 5.3803(2) \\
\em{V} (\AA$^{3}$)&      & 217.85(1) & 201.47    & 191.49\\
La/Ca             &\em{x}& 0.0157(4) & \\
                  &\em{z}&-0.0035(3) & \\
O(1)              &\em{x}&-0.0061(3) & \\
                  &\em{z}& 0.4306(2) & \\
O(2)              &\em{x}& 0.718(4)  & 0.756(1)  & 0.790(3)\\
                  &\em{y}&-0.032(1)  & \\
                  &\em{z}& 0.280(3)  & \\
R$_{WP}\;\%$      &      &  1.60     & 3.59 & 4.37\\
$\chi^{2}$        &      &  0.2      & 1.4  & 1.85\\
\noalign{\smallskip}\hline
\end{tabular}
\end{table}

\subsection{La$_{0.5}$Ca$_{0.5}$MnO$_{3}$}\label {50}

The structural refinement of La$_{0.5}$Ca$_{0.5}$MnO$_{3}$
diffraction pattern as a function of applied pressure shows that
the samples are metrically orthorhombic in the low pressure region
and the structure belongs within the P$nma$ space group up to 18
GPa. Above this pressure new peaks appear that cannot be indexed
within the P$nma$ space group. Similar to the case of x=0.25
sample, the high pressure phase turns out to be tetragonal with
I4/$mcm$ space group. The transition is well evident from the
evolution of lattice parameters (Fig.~\ref{fig:06n}(0.5-a)). The
two phases, P$nma$ and I$_4/mcm$, coexist in a reduced region
around the transition. Before the transition, we do not observe
the sluggish evolution seen for the x=0.25 sample and also the
quality of the structural refinement ($\chi^2$) remains good in
the whole range allowing to exclude the presence of other phase
transitions between the orthorhombic-to-tetragonal one.

Fitting the V(P) curve with the B-M equation\cite{Birch86} in the
P$nma$ (I$4/mcm$) phase results in a bulk modulus $B_{0}$=172 GPa
($B_{0}$=253 GPa) and its pressure derivative $B'$=5.7 ($B'$=6.4)
(table~\ref{tab:1}). As for x=0.25 sample, the bulk modulus for
x=0.5 sample is larger in the I4/$mcm$ phase pointing out the
reinforcement of stiffness through the orthorhombic-to-tetragonal
transition.

La$_{0.5}$Ca$_{0.5}$MnO$_3$ has three independent Mn-O distances
in the orthorhombic phase reported in figure~\ref{fig:06n}(0.5-c).
At low pressures, the MnO$_6$ octahedron has six almost equal
distances. As the pressure increases above $\approx 6$ GPa we
observe the progressive splitting of the three Mn-O distances till
about 10 GPa where the JT effect reaches the maximum distortion
value of about 0.07~\AA. At still higher pressures, the Mn-O
distances approach each other reducing the JT distortion
(figure~\ref{fig:06n}(0.5-c)) just before the orthorhombic to
tetragonal transition. Nevertheless, in contrast to the x=0.25
sample, structural refinement gives two distinct Mn-O distances in
the I$4/mcm$ phase pointing out the stability of the coherent JT
effect in that phase.

The $\rm{Os}_{\parallel}$ and $\rm{Os}_{\perp}$
(figure~\ref{fig:06n}(0.5-d)) are negative and decreases
(increases in absolute value) raising the pressure toward the
orthorhombic to tetragonal transition. Differently from x=0 and
x=0.25 samples $\rm{Os}_{\parallel}$ and $\rm{Os}_{\perp}$ present
very similar values as a function of pressure in the orthorhombic
phase. The P$nma$ to I$4/mcm$ phase transition provokes the drop
of $\rm{Os}_{\parallel}$ and the change of sign of
$\rm{Os}_{\perp}$ which absolute values increases steeply from
$\sim 1$\% to about 3-4 \%.

\begin{figure}[hbt]
\begin{center}
\includegraphics[width=7.3cm]{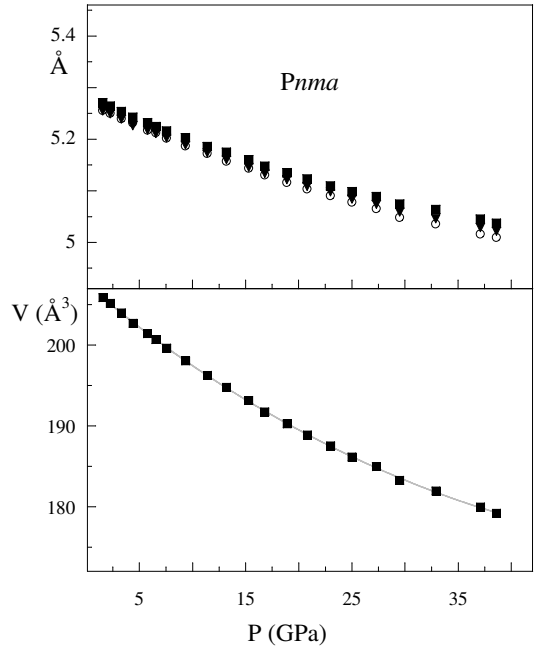}
\end{center} \caption{Lattice parameters (upper panel) and volume (lover panel) for CaMnO$_3$ sample.
The continuous grey line represent the fitting of V(P) with a B-M
equation of state.}
\label{fig:07n}       % Give a unique label
\end{figure}

\subsection{ La$_{0.33}$Ca$_{0.67}$MnO$_{3}$}\label{67}

The structural evolution of La$_{0.33}$Ca$_{0.67}$MnO$_{3}$ as a
function of the applied pressure (figure~\ref{fig:06n}(0.67-a,b))
closely resemble that of La$_{0.5}$Ca$_{0.5}$MnO$_{3}$ with a
first order P$nma$ to I$4/mcm$ phase transition. However, this
transition occurs at a lower pressure, that is around 13 GPa, for
x=0.67 sample. The two phases coexist till about 19 GPa. As for
the x=0.5 sample, we do not have evidences of a pre-transition
effect below the orthorhombic to tetragonal phase transition.
Fitting of the V(P) curve with a third order Birch-Murnaghan
equation\cite{Birch86} gives the bulk modulus $B_{0}$=210 GPa and
its pressure derivative at zero pressure $B'$=2 (table~\ref
{tab:1}) in the P$nma$ phase. In the I$4/mcm$ phase, the bulk
modulus slightly decreases to $B_{0}\approx 190$ GPa. The
refinement of the atomic positions point to three different Mn-O
distances in the P$nma$ phase (figure~\ref{fig:06n}(0.67-c)),
distinguishable over the statistical uncertainty raising the
pressure above $\approx$ 3GPa. The maximum differences between the
three Mn-O distances is achieved at around 8-10 GPa where the JT
distortion reach about 0.08\AA. At higher pressures, the
differences become progressively smaller. In the region of
coexistence of orthorhombic and tetragonal phases it is difficult
to achieve a good refinement of atomic positions, however two
distinct Mn-O distances are observed in the high pressure
tetragonal phase suggesting the stability of the coherent JT
effect in that high pressure phase.

As for the other samples the $\rm{Os}_{\parallel}$ and
$\rm{Os}_{\perp}$ (figure~\ref{fig:06n}(0.67-d)) are negative and
decreases (increases in absolute value) raising the pressure
toward the orthorhombic to tetragonal transition. We notice that
$\rm{Os}_{\parallel}$ and $\rm{Os}_{\perp}$ are similar in the
P$nma$ phase, and have similar trend, as already found in x=0.5
sample and differently from x=0 and x=0.25 samples. The
orthorhombic to tetragonal phase transition is characterized by a
a large discontinuity in $\rm{Os}_{\perp}$ increasing from $\sim
-1\%$ to more than 4\%.

\begin{table}[h]
\caption{\label{t:Ca67} Refined structural parameters from
refinement of La$_{0.33}$Ca$_{0.67}$MnO$_{3}$ XRD patterns at
selected pressures in the $Pnma$ and $I_{4}/mcm$ space groups.
Numbers in parentheses are statistical errors at the last
significant digit. The atomic positions in the $Pnma$ are: Mn (0.,
0., 0.5); La and O(1): (\emph{x}, 0.25, \emph{z}); O(2)
(\emph{x},\emph{y},\emph{z}). In the $I_{4}/mcm$ space group the
atomic positions are: Mn (0., 0., 0.); La/Ca and O(1)
 (0., 0., 0.25) and O(2) (\emph{x}, \emph{x}+0.5, 0.). }
\begin{tabular}{llll}
\hline\noalign{\smallskip}
    &    &3.5~GPa   & 19.1 GPa \\
    &    & (P$nma$) & (I$_{4}/mcm$) \\
\noalign{\smallskip}\hline\noalign{\smallskip}
\em{a}(\AA)      &      & 5.3386(4) & 5.1444(4) \\
\em{b}(\AA)      &      & 7.5273(5) & 7.6152(7) \\
\em{c}(\AA)      &      & 5.3698(4) & \\
\em{V}(\AA$^{3}$)&      & 215.79(1) & 201.53    \\
La/Ca            &\em{x}& 0.089(6)  & \\
                 &\em{z}& 0.0145(6) & \\
O(1)             &\em{x}& 0.576(6)  & \\
                 &\em{z}& 0.002(2)  & \\
O(2)             &\em{x}& 0.275(5)  & 0.286(1)\\
                 &\em{y}& 0.0217(1) & \\
                 &\em{z}& 0.693(3)  & \\
R$_{WP}\;\%$     &      &  2.40     & 2.23\\
$\chi^{2}$       &      &  0.3      & 0.9\\
\noalign{\smallskip}\hline
\end{tabular}
\end{table}

\subsection{CaMnO$_{3}$}

In CaMnO$_3$, unlike all the other compounds along in this series,
no phase transition was detected up to 45 GPa
(figure~\ref{fig:07n}). As described before, there are only
Mn$^{4+}$ manganese ions in this compound, which adopt a low spin
configuration and no Jahn-Teller distortion. The stability of the
CaMnO$_3$ structure under applied pressure demonstrates that
Mn$^{3+}$ doping is essential in stabilizing the high pressure
tetragonal phase.

The V(P) curve in CaMnO$_3$ sample has been fitted with a third
order Birch-Murnagan equation giving $B_0$=171 GPa and $B'$ = 5.9.
In this sample, $B_{0}$ slightly decreases with respect to the
value found in x=0.67. This decreasing trend contrasts the
increasing of $B_0$ as a function of x found in the other samples.
This effect is partially due to the neglect of fourth order terms
in the B-M fitting that, in addition, gives a quite large $B'$
that contrasts with the isotropic compressibility of this
structure. The fourth order fit (table~\ref{tab:1}) gives indeed a
larger $B_0\sim$195 GPa and a lower $B'\sim$2.9.

\begin{figure}[h]
\begin{center}
\includegraphics[width=7.3cm]{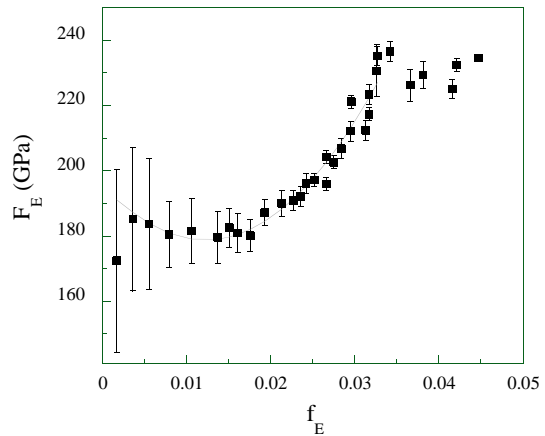}
\end{center}
\caption{\label{fig:08n} Normalized pressure
F$_{E}$=P/3f$_{E}$(1+2f$_{E}$) versus the Eulerian strain
f$_{E}$=$\frac{(V/V_{0})^{2/3}-1}{2}$ for x=0.25 sample. $V_{0}$
is the cell volume at ambient conditions.}
       % Give a unique label
\end{figure}

\section{discussion}

The plots in figure \ref{fig:06n} describe the structural phase
diagram of La$_{1-x}$Ca$_{x}$MnO$_3$ as a function of composition
and applied pressure. The extreme compounds remain orthorhombic in
the whole pressure range investigated. Nevertheless the LaMnO$_3$
depicts a complex evolution of the cell edges and a P$nma$ to
I$mma$ phase transition that are absent in the CaMnO$_3$ sample,
where the only effect of pressure is the isotropic squeezing of
the unit cell. This difference must be related to the JT nature of
Mn$^{3+}$ ions, in x=0 samples, opposite to the undistorted
MnO$_6$ octahedra in x=1 samples.

All the doped compounds, at low pressure, are in the orthorhombic
phase and evolve toward a tetragonal one on raising the pressure.
This phase transition is closely related to the Ca doping: the
onset of orthorhombic to tetragonal transition moves to a lower
pressure as the Ca content increases. Moreover, the nature of this
transition is different in the hole doped  ($ x < 0.5$) and in the
electron doped ($x\geq 0.5$) composition regions: in x=0.25 the
transition appears smooth and the presence of intermediate
parasitic phases can be hypothesized on the basis of the reduced
refinement accuracy (larger $\chi^2$). On the contrary the x=0.5
and x=0.67 samples depict first order P$nma$ to I$4/mcm$
transition.

The increasing of Ca content has a clear effect on the energetic
cost for the orthorhombic to tetragonal transition ($\Delta E =
P\Delta V$). For x=0.75 the P$nma$-I$4/mcm$ phase transition is
very smooth and we found $\Delta V$ =V(tetra)-V(ortho)=-0.53
\AA$^{3}$ at P = 26.4 GPa, giving to $\Delta E = -8.4$ KJ/mole.
For x=0.50 $\Delta V =1.07 \AA^{3}$ at P=19.5 GPa corresponding to
$\Delta E =-12.56 $ KJ/mole and, for x=0.33 $\Delta V = 2.33
\AA^{3}$ at P=17.33 GPa, corresponding to a $\Delta E = -24.33$
KJ/mole. This means that on increasing the Ca doping the
tetragonal phase needs more energy to be stable and this is
confirmed by the progressive widening of the coexistence region of
the two phases in figure \ref{fig:06n}. To explain this finding,
we notice that the P$nma$ structure responds to the hydrostatic
compression not only by squeezing the MnO$_6$ octahedra but also
by acting on their relative orientation, i.e.
 changing the Mn-O-Mn bond angle, that is less expensive for
the energetic balance. In the I$mma$ and I$4/mcm$ phases, the
tilting mode changes and the octahedra are aligned along one
direction: along the [101] direction in the I$mma$ phase (figure
\ref{fig:03n}b) and along the b axis in the I$4/mcm$ phase (figure
\ref{fig:03n}c). Moreover, the coordination for the cation (La/Ca)
changes from 8 (P$nma$) to 10-12 in the I$4/mcm$-I$mma$ phase.
These changes make the structure more rigid as confirmed by the
overall increasing of the lattice stiffness across the
P$nma$-I$4/mcm$ phase transition (table~\ref{tab:1}).

All the doped compounds, in a range of the high pressure region,
exhibit an evident splitting of Mn-O distances signalling the
enhancement of coherent JT distortions. The analogous trend  has
been already reported for x=0.25 sample~\cite{MeneghiniHP}.
Moreover in x=0.5 and x=0.67 samples two Mn-O distances remains
well distinct even in the high pressure tetragonal phase. These
findings, together with previous high pressure
resistivity~\cite{MeneghiniHP}, Raman~\cite{Postorino01} and infra
red~\cite{Congeduti01} spectroscopy results, demonstrate that, at
high pressures, a mechanism competes with the pressure induced
reduction of the distortions and charge delocalization, as
observed at the low pressures~\cite{Laukhin97,Hwang95}. This
mechanism is largely unexpected on the basis of a model involving
only DE and e-p interactions. In fact squeezing the structure
raises e$_g$ electron hopping integral; thus, increased pressure
is expected to enhance the charge mobility. In addition, since the
polaron stability is inversely proportional to the charge
mobility, the squeezing of the structure is expected to bring the
system in a less distorted-metallic phase, in contrast to the
experimental result, at least in some regions of the high
pressure-concentration phase diagram.

To explain these findings, other mechanisms must be taken into
account, and these could be the cooperative JT effect and the SE
coupling. The stability of JT distortion is related to the
difference between the energy gained by the on-site distortion and
the energy spent by the lattice to bring about such a distortion.
In case of cooperative JT effect, the ordered orbital allow for
alternating elongated and compressed Mn-O bonds, reducing the net
lattice distortion, thus making the JT polaron energetically more
stable. This effect has been invoked \cite{Tokura96} to explain
the pressure induced stabilization of a low temperature CO phase
to the detriment of FM phase in (Nd,Sm)$_{0.5}$Sr$_{0.5}$MnO$_3$
perovskites. On the other side the SE coupling, being fundamental
in stabilizing the CO phases in x~$\geq 0.5$ compounds, must also
play a role. Theoretical simulations have indeed suggested the
possibility of FM to CO phase transition as a function of the SE
coupling strength\cite{Yunoki00,Fratini01}.

We have studied structural the high pressure phase diagram of
La$_{1-x}$Ca$_x$MnO$_3$ perovskites. Our findings detail the
peculiar structural evolution of sample structure as a function of
pressure and composition and suggest that squeezing the structure
under hydrostatic pressure has a profound effect on the relative
strength of the involved interactions, namely DE, SE and charge to
lattice coupling.

% BibTeX users please use
% BibTeX users please use
%\bibliography{manuscript}

\begin{thebibliography}{99}
\expandafter\ifx\csname
natexlab\endcsname\relax\def\natexlab#1{#1}\fi
\expandafter\ifx\csname bibnamefont\endcsname\relax
  \def\bibnamefont#1{#1}\fi
\expandafter\ifx\csname bibfnamefont\endcsname\relax
  \def\bibfnamefont#1{#1}\fi
\expandafter\ifx\csname citenamefont\endcsname\relax
  \def\citenamefont#1{#1}\fi
\expandafter\ifx\csname url\endcsname\relax
  \def\url#1{\texttt{#1}}\fi
\expandafter\ifx\csname
urlprefix\endcsname\relax\def\urlprefix{URL }\fi
\providecommand{\bibinfo}[2]{#2}
\providecommand{\eprint}[2][]{\url{#2}}

\bibitem[{\citenamefont{Zener}(1951)}]{Zener51}
\bibinfo{author}{\bibfnamefont{C.}~\bibnamefont{Zener}},
  \bibinfo{journal}{Phys. Rev.} \textbf{\bibinfo{volume}{82}},
  \bibinfo{pages}{403} (\bibinfo{year}{1951}).

\bibitem[{\citenamefont{Anderson and Hasegawa}(1955)}]{Anderson55}
\bibinfo{author}{\bibfnamefont{P.}~\bibnamefont{Anderson}} \bibnamefont{and}
  \bibinfo{author}{\bibfnamefont{H.}~\bibnamefont{Hasegawa}},
  \bibinfo{journal}{Phys.Rev.} \textbf{\bibinfo{volume}{100}},
  \bibinfo{pages}{675} (\bibinfo{year}{1955}).

\bibitem[{\citenamefont{de~Gennes}(1960)}]{Gennes60}
\bibinfo{author}{\bibfnamefont{P.}~\bibnamefont{de~Gennes}},
  \bibinfo{journal}{Phys. Rev.} \textbf{\bibinfo{volume}{118}},
  \bibinfo{pages}{141} (\bibinfo{year}{1960}).

\bibitem[{\citenamefont{Millis et~al.}(1995)\citenamefont{Millis, Littlewood,
  and Shraiman}}]{Millis95}
\bibinfo{author}{\bibfnamefont{A.~J.} \bibnamefont{Millis}},
  \bibinfo{author}{\bibfnamefont{P.~B.}~\bibnamefont{Littlewood}},
  \bibnamefont{and} \bibinfo{author}{\bibfnamefont{B.~I.}~\bibnamefont{Shraiman}},
  \bibinfo{journal}{Phys. Rev. Lett.} \textbf{\bibinfo{volume}{74}},
  \bibinfo{pages}{5144} (\bibinfo{year}{1995}).

\bibitem[{\citenamefont{Millis et~al.}(1996)\citenamefont{Millis, Shraiman, and
  Mueller}}]{Millis96}
\bibinfo{author}{\bibfnamefont{A.~J.} \bibnamefont{Millis}},
  \bibinfo{author}{\bibfnamefont{B.~I.} \bibnamefont{Shraiman}},
  \bibnamefont{and} \bibinfo{author}{\bibfnamefont{R.}~\bibnamefont{Mueller}},
  \bibinfo{journal}{Phys. Rev. Lett.} \textbf{\bibinfo{volume}{77}},
  \bibinfo{pages}{175} (\bibinfo{year}{1996}).

\bibitem[{\citenamefont{Schiffer et~al.}(1995)\citenamefont{Schiffer, Ramirez,
  Bao, and Cheong}}]{Schiffer95}
\bibinfo{author}{\bibfnamefont{P.}~\bibnamefont{Schiffer}},
  \bibinfo{author}{\bibfnamefont{A.~P.} \bibnamefont{Ramirez}},
  \bibinfo{author}{\bibfnamefont{W.}~\bibnamefont{Bao}}, \bibnamefont{and}
  \bibinfo{author}{\bibfnamefont{S.-W.} \bibnamefont{Cheong}},
  \bibinfo{journal}{Phys. Rev. Lett.} \textbf{\bibinfo{volume}{75}},
  \bibinfo{pages}{3336} (\bibinfo{year}{1995}).

\bibitem[{\citenamefont{Fern\'andez-D\'iaz and et~al.}(1999)}]{Fernandez99}
\bibinfo{author}{\bibfnamefont{M.}~\bibnamefont{Fern\'andez-D\'iaz}}
  \bibnamefont{and} \bibinfo{author}{\bibnamefont{et~al.}},
  \bibinfo{journal}{Phys. Rev. B} \textbf{\bibinfo{volume}{59}},
  \bibinfo{pages}{1277} (\bibinfo{year}{1999}).

\bibitem[{\citenamefont{Ohsawa and Inoue}(2001)}]{Ohsawa01}
\bibinfo{author}{\bibfnamefont{T.}~\bibnamefont{Ohsawa}} \bibnamefont{and}
  \bibinfo{author}{\bibfnamefont{J.}~\bibnamefont{Inoue}},
  \bibinfo{journal}{Phys. Rev. B} \textbf{\bibinfo{volume}{65}},
  \bibinfo{pages}{14401} (\bibinfo{year}{2001}).

\bibitem[{\citenamefont{Booth et~al.}(1998)\citenamefont{Booth, Bridges, Kwei,
  Laurence, Cornelius, and Neumeier}}]{Booth98}
\bibinfo{author}{\bibfnamefont{C.~H.} \bibnamefont{Booth}},
  \bibinfo{author}{\bibfnamefont{F.}~\bibnamefont{Bridges}},
  \bibinfo{author}{\bibfnamefont{G.~H.} \bibnamefont{Kwei}},
  \bibinfo{author}{\bibfnamefont{J.~M.} \bibnamefont{Lawrence}},
  \bibinfo{author}{\bibfnamefont{A.~L.} \bibnamefont{Cornelius}},
  \bibnamefont{and} \bibinfo{author}{\bibfnamefont{J.~J.}
  \bibnamefont{Neumeier}}, \bibinfo{journal}{Phys. Rev. B}
  \textbf{\bibinfo{volume}{57}}, \bibinfo{pages}{10440} (\bibinfo{year}{1998}).

\bibitem[{\citenamefont{Radaelli
  et~al.}(1996{\natexlab{a}})\citenamefont{Radaelli, Cox, Marezio, Cheong,
  Shiffer, and Ramirez}}]{Radaelli96}
\bibinfo{author}{\bibfnamefont{P.~G.} \bibnamefont{Radaelli}},
  \bibinfo{author}{\bibfnamefont{D.~E.} \bibnamefont{Cox}},
  \bibinfo{author}{\bibfnamefont{M.}~\bibnamefont{Marezio}},
  \bibinfo{author}{\bibfnamefont{S.~W.} \bibnamefont{Cheong}},
  \bibinfo{author}{\bibfnamefont{P.~E.} \bibnamefont{Shiffer}},
  \bibnamefont{and} \bibinfo{author}{\bibfnamefont{A.~P.}
  \bibnamefont{Ramirez}}, \bibinfo{journal}{Phys. Rev. Lett.}
  \textbf{\bibinfo{volume}{75}}, \bibinfo{pages}{4488}
  (\bibinfo{year}{1995}{\natexlab{a}}).

\bibitem[{\citenamefont{Louca and Egami}(1999)}]{Louca99}
\bibinfo{author}{\bibfnamefont{D.}~\bibnamefont{Louca}} \bibnamefont{and}
  \bibinfo{author}{\bibfnamefont{T.}~\bibnamefont{Egami}},
  \bibinfo{journal}{Phys. Rev. B} \textbf{\bibinfo{volume}{59}},
  \bibinfo{pages}{6193} (\bibinfo{year}{1999}).

\bibitem[{\citenamefont{Louca et~al.}(2001)\citenamefont{Louca, Egami, Dmowski,
  and Mitchell}}]{Louca01}
\bibinfo{author}{\bibfnamefont{D.}~\bibnamefont{Louca}},
  \bibinfo{author}{\bibfnamefont{T.}~\bibnamefont{Egami}},
  \bibinfo{author}{\bibfnamefont{W.}~\bibnamefont{Dmowski}}, \bibnamefont{and}
  \bibinfo{author}{\bibfnamefont{J.~F.} \bibnamefont{Mitchell}},
  \bibinfo{journal}{Phys. Rev. B} \textbf{\bibinfo{volume}{64}},
  \bibinfo{pages}{180403} (\bibinfo{year}{2001}).

\bibitem[{\citenamefont{Lanzara et~al.}(1998)\citenamefont{Lanzara, Saini,
  Brunelli, Natali, Bianconi, Radaelli, and Cheong}}]{Lanzara98}
\bibinfo{author}{\bibfnamefont{A.}~\bibnamefont{Lanzara}},
  \bibinfo{author}{\bibfnamefont{N.~L.} \bibnamefont{Saini}},
  \bibinfo{author}{\bibfnamefont{M.}~\bibnamefont{Brunelli}},
  \bibinfo{author}{\bibfnamefont{F.}~\bibnamefont{Natali}},
  \bibinfo{author}{\bibfnamefont{A.}~\bibnamefont{Bianconi}},
  \bibinfo{author}{\bibfnamefont{P.~G.} \bibnamefont{Radaelli}},
  \bibnamefont{and} \bibinfo{author}{\bibfnamefont{S.~W.}
  \bibnamefont{Cheong}}, \bibinfo{journal}{Phys. Rev. Lett.}
  \textbf{\bibinfo{volume}{81}}, \bibinfo{pages}{878} (\bibinfo{year}{1998}).

\bibitem[{\citenamefont{Meneghini et~al.}(2002)\citenamefont{Meneghini,
  Castellano, Mobilio, Kumar, Ray, and Sarma}}]{Meneghini02}
\bibinfo{author}{\bibfnamefont{C.}~\bibnamefont{Meneghini}},
  \bibinfo{author}{\bibfnamefont{C.}~\bibnamefont{Castellano}},
  \bibinfo{author}{\bibfnamefont{S.}~\bibnamefont{Mobilio}},
  \bibinfo{author}{\bibfnamefont{A.}~\bibnamefont{Kumar}},
  \bibinfo{author}{\bibfnamefont{S.}~\bibnamefont{Ray}}, \bibnamefont{and}
  \bibinfo{author}{\bibfnamefont{D.~D.} \bibnamefont{Sarma}},
  \bibinfo{journal}{J. Phys. Cond. Mat.} \textbf{\bibinfo{volume}{14}},
  \bibinfo{pages}{1967} (\bibinfo{year}{2002}).

\bibitem[{\citenamefont{Sub\'{i}as et~al.}(2002)\citenamefont{Sub\'{i}as,
  Garc\'{i}a, Blasco, S\'{a}nchez, and Proietti}}]{Subias02}
\bibinfo{author}{\bibfnamefont{G.}~\bibnamefont{Sub\'{i}as}},
  \bibinfo{author}{\bibfnamefont{J.}~\bibnamefont{Garc\'{i}a}},
  \bibinfo{author}{\bibfnamefont{J.}~\bibnamefont{Blasco}},
  \bibinfo{author}{\bibfnamefont{M.~C.} \bibnamefont{S\'{a}nchez}},
  \bibnamefont{and} \bibinfo{author}{\bibfnamefont{M.~G.}
  \bibnamefont{Proietti}}, \bibinfo{journal}{J. Phys.: Cond. Mat.}
  \textbf{\bibinfo{volume}{14}}, \bibinfo{pages}{5017} (\bibinfo{year}{2002}).

\bibitem[{\citenamefont{Bardelli et~al.}(2003)\citenamefont{Bardelli,
  Meneghini, Mobilio, Castellano, and Dediu}}]{Bardelli02}
\bibinfo{author}{\bibfnamefont{F.}~\bibnamefont{Bardelli}},
  \bibinfo{author}{\bibfnamefont{C.}~\bibnamefont{Meneghini}},
  \bibinfo{author}{\bibfnamefont{S.}~\bibnamefont{Mobilio}},
  \bibinfo{author}{\bibfnamefont{C.}~\bibnamefont{Castellano}},
  \bibnamefont{and} \bibinfo{author}{\bibfnamefont{V.}~\bibnamefont{Dediu}},
  \bibinfo{journal}{Nucl. Inst. and Met. B} \textbf{\bibinfo{volume}{200}},
  \bibinfo{pages}{226} (\bibinfo{year}{2003}).

\bibitem[{\citenamefont{Hwang et~al.}(1995{\natexlab{a}})\citenamefont{Hwang,
  Cheong, Radaelli, Marezio, and Batlogg}}]{Hwang95b}
\bibinfo{author}{\bibfnamefont{H.~Y.} \bibnamefont{Hwang}},
  \bibinfo{author}{\bibfnamefont{S.-W.} \bibnamefont{Cheong}},
  \bibinfo{author}{\bibfnamefont{P.~G.} \bibnamefont{Radaelli}},
  \bibinfo{author}{\bibfnamefont{M.}~\bibnamefont{Marezio}}, \bibnamefont{and}
  \bibinfo{author}{\bibfnamefont{B.}~\bibnamefont{Batlogg}},
  \bibinfo{journal}{Phys. Rev. Lett.} \textbf{\bibinfo{volume}{75}},
  \bibinfo{pages}{914} (\bibinfo{year}{1995}{\natexlab{a}}).

\bibitem[{\citenamefont{Yoshizawa et~al.}(1997)\citenamefont{Yoshizawa,
  Kajimoto, Kawano, Tomioka, and tokura}}]{Yoshizawa98}
\bibinfo{author}{\bibfnamefont{H.}~\bibnamefont{Yoshizawa}},
  \bibinfo{author}{\bibfnamefont{R.}~\bibnamefont{Kajimoto}},
  \bibinfo{author}{\bibfnamefont{H.}~\bibnamefont{Kawano}},
  \bibinfo{author}{\bibfnamefont{Y.}~\bibnamefont{Tomioka}}, \bibnamefont{and}
  \bibinfo{author}{\bibfnamefont{Y.}~\bibnamefont{Tokura}},
  \bibinfo{journal}{Phys. Rev B} \textbf{\bibinfo{volume}{55}},
  \bibinfo{pages}{2729} (\bibinfo{year}{1997}).

\bibitem[{\citenamefont{Hwang et~al.}(1995{\natexlab{b}})\citenamefont{Hwang,
  Palstra, Cheong, and Batlogg}}]{Hwang95}
\bibinfo{author}{\bibfnamefont{H.~Y.} \bibnamefont{Hwang}},
  \bibinfo{author}{\bibfnamefont{T.~T.~M.} \bibnamefont{Palstra}},
  \bibinfo{author}{\bibfnamefont{S.~W.} \bibnamefont{Cheong}},
  \bibnamefont{and} \bibinfo{author}{\bibfnamefont{B.}~\bibnamefont{Batlogg}},
  \bibinfo{journal}{Phys. Rev. B} \textbf{\bibinfo{volume}{52}},
  \bibinfo{pages}{15046} (\bibinfo{year}{1995}{\natexlab{b}}).

\bibitem[{\citenamefont{Nossov et~al.}(1998)\citenamefont{Nossov, Pierre,
  Beille, Vassiliev, and Slobodin}}]{Nossov98}
\bibinfo{author}{\bibfnamefont{A.}~\bibnamefont{Nossov}},
  \bibinfo{author}{\bibfnamefont{J.}~\bibnamefont{Pierre}},
  \bibinfo{author}{\bibfnamefont{J.}~\bibnamefont{Beille}},
  \bibinfo{author}{\bibfnamefont{V.}~\bibnamefont{Vassiliev}},
  \bibnamefont{and} \bibinfo{author}{\bibfnamefont{B.}~\bibnamefont{Slobodin}},
  \bibinfo{journal}{Eur. Phys. J. B} \textbf{\bibinfo{volume}{6}},
  \bibinfo{pages}{467} (\bibinfo{year}{1998}).

\bibitem[{\citenamefont{Neumeier et~al.}(1995)\citenamefont{Neumeier, Hundley,
  Thompson, and Heffner}}]{Neumeier95}
\bibinfo{author}{\bibfnamefont{J.~J.}~\bibnamefont{Neumeier}},
  \bibinfo{author}{\bibfnamefont{M.~F.} \bibnamefont{Hundley}},
  \bibinfo{author}{\bibfnamefont{J.~D.} \bibnamefont{Thompson}},
  \bibnamefont{and} \bibinfo{author}{\bibfnamefont{R.~H.}
  \bibnamefont{Heffner}}, \bibinfo{journal}{Phys. Rev. B}
  \textbf{\bibinfo{volume}{52}}, \bibinfo{pages}{7006} (\bibinfo{year}{1995}).

\bibitem[{\citenamefont{Tokura et~al.}(1996)\citenamefont{Tokura, Kuwahara,
  Morimoto, Tomida, and Asamitsu}}]{Tokura96}
\bibinfo{author}{\bibfnamefont{Y.}~\bibnamefont{Tokura}},
  \bibinfo{author}{\bibfnamefont{H.}~\bibnamefont{Kuwahara}},
  \bibinfo{author}{\bibfnamefont{Y.}~\bibnamefont{Moritomo}},
  \bibinfo{author}{\bibfnamefont{Y.}~\bibnamefont{Tomioka}}, \bibnamefont{and}
  \bibinfo{author}{\bibfnamefont{A.}~\bibnamefont{Asamitsu}},
  \bibinfo{journal}{Phys. Rev. Lett.} \textbf{\bibinfo{volume}{76}},
  \bibinfo{pages}{3184} (\bibinfo{year}{1996}).

\bibitem[{\citenamefont{Laukhin et~al.}(1997)\citenamefont{Laukhin,
  Fontcuberta, Garcia-Munoz, and Obrados}}]{Laukhin97}
\bibinfo{author}{\bibfnamefont{V.}~\bibnamefont{Laukhin}},
  \bibinfo{author}{\bibfnamefont{J.}~\bibnamefont{Fontcuberta}},
  \bibinfo{author}{\bibfnamefont{J.~L.} \bibnamefont{Garcia-Munoz}},
  \bibnamefont{and} \bibinfo{author}{\bibfnamefont{X.}~\bibnamefont{Obradors}},
  \bibinfo{journal}{Phys. Rev. B} \textbf{\bibinfo{volume}{56}},
  \bibinfo{pages}{10009} (\bibinfo{year}{1997}).

\bibitem[{\citenamefont{Senis et~al.}(1998)\citenamefont{Senis, Laukhin,
  Martinez, Fontcuberta, Obrados, Arsenov, and Mukovskii}}]{Senis98}
\bibinfo{author}{\bibfnamefont{R.}~\bibnamefont{Senis}},
  \bibinfo{author}{\bibfnamefont{V.}~\bibnamefont{Laukhin}},
  \bibinfo{author}{\bibfnamefont{B.}~\bibnamefont{Martinez}},
  \bibinfo{author}{\bibfnamefont{J.}~\bibnamefont{Fontcuberta}},
  \bibinfo{author}{\bibfnamefont{X.}~\bibnamefont{Obradors}},
  \bibinfo{author}{\bibfnamefont{A.~A.} \bibnamefont{Arsenov}},
  \bibnamefont{and} \bibinfo{author}{\bibfnamefont{Y.~M.}
  \bibnamefont{Mukovskii}}, \bibinfo{journal}{Phys. Rev. B}
  \textbf{\bibinfo{volume}{57}}, \bibinfo{pages}{14680} (\bibinfo{year}{1998}).

\bibitem[{\citenamefont{Congeduti
  et~al.}(2001{\natexlab{a}})\citenamefont{Congeduti, Postorino, Dore, Nucara,
  Lupi, Mercone, Calvani, and Sarma}}]{Congeduti01}
\bibinfo{author}{\bibfnamefont{A.}~\bibnamefont{Congeduti}},
  \bibinfo{author}{\bibfnamefont{P.}~\bibnamefont{Postorino}},
  \bibinfo{author}{\bibfnamefont{P.}~\bibnamefont{Dore}},
  \bibinfo{author}{\bibfnamefont{A.}~\bibnamefont{Nucara}},
  \bibinfo{author}{\bibfnamefont{S.}~\bibnamefont{Lupi}},
  \bibinfo{author}{\bibfnamefont{S.}~\bibnamefont{Mercone}},
  \bibinfo{author}{\bibfnamefont{P.}~\bibnamefont{Calvani}},
  \bibinfo{author}{\bibfnamefont{A.}~\bibnamefont{Kumar}}, \bibnamefont{and}
  \bibinfo{author}{\bibfnamefont{D.~D.} \bibnamefont{Sarma}},
  \bibinfo{journal}{Phys. Rev. B} \textbf{\bibinfo{volume}{63}},
  \bibinfo{pages}{184410} (\bibinfo{year}{2001}{\natexlab{a}}).

\bibitem[{\citenamefont{Congeduti
  et~al.}(2001{\natexlab{b}})\citenamefont{Congeduti, Postorino, Caramagno,
  Nardone, Kumar, and Sarma}}]{Postorino01}
\bibinfo{author}{\bibfnamefont{A.}~\bibnamefont{Congeduti}},
  \bibinfo{author}{\bibfnamefont{P.}~\bibnamefont{Postorino}},
  \bibinfo{author}{\bibfnamefont{E.}~\bibnamefont{Caramagno}},
  \bibinfo{author}{\bibfnamefont{M.}~\bibnamefont{Nardone}},
  \bibinfo{author}{\bibfnamefont{A.}~\bibnamefont{Kumar}}, \bibnamefont{and}
  \bibinfo{author}{\bibfnamefont{D.~D.} \bibnamefont{Sarma}},
  \bibinfo{journal}{Phys. Rev. Lett.} \textbf{\bibinfo{volume}{86}},
  \bibinfo{pages}{1251} (\bibinfo{year}{2001}{\natexlab{b}}).

\bibitem[{\citenamefont{Meneghini et~al.}(2001)\citenamefont{Meneghini, Levy,
  Mobilio, Ortolani, Reguero, Kumar, and Sarma}}]{MeneghiniHP}
\bibinfo{author}{\bibfnamefont{C.}~\bibnamefont{Meneghini}},
  \bibinfo{author}{\bibfnamefont{D.}~\bibnamefont{Levy}},
  \bibinfo{author}{\bibfnamefont{S.}~\bibnamefont{Mobilio}},
  \bibinfo{author}{\bibfnamefont{M.}~\bibnamefont{Ortolani}},
  \bibinfo{author}{\bibfnamefont{M.} \bibnamefont{Nunez-Reguero}},
  \bibinfo{author}{\bibfnamefont{A.}~\bibnamefont{Kumar}}, \bibnamefont{and}
  \bibinfo{author}{\bibfnamefont{D.~D.} \bibnamefont{Sarma}},
  \bibinfo{journal}{Phys. Rev. B} \textbf{\bibinfo{volume}{65}},
  \bibinfo{pages}{12111} (\bibinfo{year}{2001}).

\bibitem[{\citenamefont{Schulze et~al.}(1998)\citenamefont{Schulze, Lienert,
  Hanfland, Lorenzen, and Zontone}}]{zonton}
\bibinfo{author}{\bibfnamefont{C.}~\bibnamefont{Schulze}},
  \bibinfo{author}{\bibfnamefont{U.}~\bibnamefont{Lienert}},
  \bibinfo{author}{\bibfnamefont{M.}~\bibnamefont{Hanfland}},
  \bibinfo{author}{\bibfnamefont{M.}~\bibnamefont{Lorenzen}}, \bibnamefont{and}
  \bibinfo{author}{\bibfnamefont{F.}~\bibnamefont{Zontone}},
  \bibinfo{journal}{J.Synchrotron Rad.} \textbf{\bibinfo{volume}{5}},
  \bibinfo{pages}{77} (\bibinfo{year}{1998}).

\bibitem[{\citenamefont{Mao et~al.}(1986)\citenamefont{Mao, Xu, and
  Bell}}]{Mao86}
\bibinfo{author}{\bibfnamefont{H.~K.} \bibnamefont{Mao}},
  \bibinfo{author}{\bibfnamefont{J.}~\bibnamefont{Xu}}, \bibnamefont{and}
  \bibinfo{author}{\bibfnamefont{P.~M.} \bibnamefont{Bell}},
  \bibinfo{journal}{J. Geophys. Res.} \textbf{\bibinfo{volume}{91}},
  \bibinfo{pages}{16491} (\bibinfo{year}{1986}).

\bibitem[{\citenamefont{Hammersley et~al.}(1996)\citenamefont{Hammersley,
  Svensson, Hanfland, Fitch, and Hauserman}}]{FIT2D}
\bibinfo{author}{\bibfnamefont{A.~P.} \bibnamefont{Hammersley}},
  \bibinfo{author}{\bibfnamefont{S.~O.} \bibnamefont{Svensson}},
  \bibinfo{author}{\bibfnamefont{M.}~\bibnamefont{Hanfland}},
  \bibinfo{author}{\bibfnamefont{A.~N.} \bibnamefont{Fitch}}, \bibnamefont{and}
  \bibinfo{author}{\bibfnamefont{D.}~\bibnamefont{Hauserman}},
  \bibinfo{journal}{High Pressure Res.} \textbf{\bibinfo{volume}{14}},
  \bibinfo{pages}{235} (\bibinfo{year}{1996}).

\bibitem[{\citenamefont{Larsson and von Dreele}(1999)}]{GSAS}
\bibinfo{author}{\bibfnamefont{A.~C.} \bibnamefont{Larsson}} \bibnamefont{and}
  \bibinfo{author}{\bibfnamefont{R.~B.} \bibnamefont{von Dreele}},
  \bibinfo{journal}{Report LAUR} \textbf{\bibinfo{volume}{86-748}},
  \bibinfo{pages}{Los Alamos National Laboratory, Los Alamos, New Mexico}
  (\bibinfo{year}{1999}).

\bibitem[{\citenamefont{Birch}(1947)}]{Birch47}
\bibinfo{author}{\bibfnamefont{F.}~\bibnamefont{Birch}},
  \bibinfo{journal}{Phys. Rev.} \textbf{\bibinfo{volume}{71}},
  \bibinfo{pages}{809} (\bibinfo{year}{1947}).

\bibitem[{\citenamefont{Loa et~al.}(2001)\citenamefont{Loa, Adler, Grzechnik,
  Syassen, Schwarz, Hanfland, Rozenberg, Gorodetsky, and Pasternak}}]{Ingo01}
\bibinfo{author}{\bibfnamefont{I.}~\bibnamefont{Loa}},
  \bibinfo{author}{\bibfnamefont{P.}~\bibnamefont{Adler}},
  \bibinfo{author}{\bibfnamefont{A.}~\bibnamefont{Grzechnik}},
  \bibinfo{author}{\bibfnamefont{K.}~\bibnamefont{Syassen}},
  \bibinfo{author}{\bibfnamefont{U.}~\bibnamefont{Schwarz}},
  \bibinfo{author}{\bibfnamefont{M.}~\bibnamefont{Hanfland}},
  \bibinfo{author}{\bibfnamefont{G.~K.} \bibnamefont{Rozenberg}},
  \bibinfo{author}{\bibfnamefont{P.}~\bibnamefont{Gorodetsky}},
  \bibnamefont{and} \bibinfo{author}{\bibfnamefont{M.~P.}
  \bibnamefont{Pasternak}}, \bibinfo{journal}{Phys. Rev. Lett.}
  \textbf{\bibinfo{volume}{87}}, \bibinfo{pages}{125501}
  (\bibinfo{year}{2001}).

\bibitem[{\citenamefont{Radaelli
  et~al.}(1996{\natexlab{b}})\citenamefont{Radaelli, Marezio, Hwang, and
  Cheong}}]{radaelli96b}
\bibinfo{author}{\bibfnamefont{P.~G.} \bibnamefont{Radaelli}},
  \bibinfo{author}{\bibfnamefont{M.}~\bibnamefont{Marezio}},
  \bibinfo{author}{\bibfnamefont{H.~Y.} \bibnamefont{Hwang}}, \bibnamefont{and}
  \bibinfo{author}{\bibfnamefont{S.-W.} \bibnamefont{Cheong}},
  \bibinfo{journal}{J. Solid State Chem.} \textbf{\bibinfo{volume}{122}},
  \bibinfo{pages}{444} (\bibinfo{year}{1996}{\natexlab{b}}).

\bibitem[{\citenamefont{Birch}(1986)}]{Birch86}
\bibinfo{author}{\bibfnamefont{F.~J.} \bibnamefont{Birch}},
  \bibinfo{journal}{J. Geophys. Res.} \textbf{\bibinfo{volume}{91}},
  \bibinfo{pages}{4949} (\bibinfo{year}{1986}).

\bibitem[{\citenamefont{Gaudart et~al.}(2001)\citenamefont{Gaudart, Carvajal,
  Aladine, Goncharenko, Medarde, Smith, and Revcolevschi}}]{Pinsard01}
\bibinfo{author}{\bibfnamefont{L.} \bibnamefont{Pinsard-Gaudart}},
  \bibinfo{author}{\bibfnamefont{J.} \bibnamefont{Rodriguez-Carvajal}},
  \bibinfo{author}{\bibfnamefont{A.} \bibnamefont{Daoud-Aladine}},
  \bibinfo{author}{\bibfnamefont{I.}~\bibnamefont{Goncharenko}},
  \bibinfo{author}{\bibfnamefont{M.}~\bibnamefont{Medarde}},
  \bibinfo{author}{\bibfnamefont{R.~I.} \bibnamefont{Smith}}, \bibnamefont{and}
  \bibinfo{author}{\bibfnamefont{A.}~\bibnamefont{Revcolevschi}},
  \bibinfo{journal}{Phys. Rev. B} \textbf{\bibinfo{volume}{64}},
  \bibinfo{pages}{64426} (\bibinfo{year}{2001}).

\bibitem[{\citenamefont{Woodward et~al.}(1998)\citenamefont{Woodward, Vogt,
  Cox, Arulraj, Rao, Karen, and Cheetham}}]{Woodward98}
\bibinfo{author}{\bibfnamefont{P.~M.} \bibnamefont{Woodward}},
  \bibinfo{author}{\bibfnamefont{T.}~\bibnamefont{Vogt}},
  \bibinfo{author}{\bibfnamefont{D.~E.} \bibnamefont{Cox}},
  \bibinfo{author}{\bibfnamefont{A.}~\bibnamefont{Arulraj}},
  \bibinfo{author}{\bibfnamefont{C.~N.~R.} \bibnamefont{Rao}},
  \bibinfo{author}{\bibfnamefont{P.}~\bibnamefont{Karen}}, \bibnamefont{and}
  \bibinfo{author}{\bibfnamefont{A.~K.} \bibnamefont{Cheetham}},
  \bibinfo{journal}{Chem. Mater.} \textbf{\bibinfo{volume}{10}},
  \bibinfo{pages}{3652} (\bibinfo{year}{1998}).

\bibitem[{\citenamefont{Glazer}(1972)}]{Glazer72}
\bibinfo{author}{\bibfnamefont{A.~M.} \bibnamefont{Glazer}},
  \bibinfo{journal}{Acta Cryst.} \textbf{\bibinfo{volume}{B28}},
  \bibinfo{pages}{3384} (\bibinfo{year}{1972}).

\bibitem[{\citenamefont{Yunoki et~al.}(2000)\citenamefont{Yunoki, Hotta, and
  Dagotto}}]{Yunoki00}
\bibinfo{author}{\bibfnamefont{S.}~\bibnamefont{Yunoki}},
  \bibinfo{author}{\bibfnamefont{T.}~\bibnamefont{Hotta}}, \bibnamefont{and}
  \bibinfo{author}{\bibfnamefont{E.}~\bibnamefont{Dagotto}},
  \bibinfo{journal}{Phys. Rev. Lett.} \textbf{\bibinfo{volume}{84}},
  \bibinfo{pages}{3714} (\bibinfo{year}{2000}).

\bibitem[{\citenamefont{Fratini et~al.}(2001)\citenamefont{Fratini, Feinberg,
  and Grilli}}]{Fratini01}
\bibinfo{author}{\bibfnamefont{S.}~\bibnamefont{Fratini}},
  \bibinfo{author}{\bibfnamefont{D.}~\bibnamefont{Feinberg}}, \bibnamefont{and}
  \bibinfo{author}{\bibfnamefont{M.}~\bibnamefont{Grilli}},
  \bibinfo{journal}{Europ. Phys. J.} p. \bibinfo{pages}{to be published}
  (\bibinfo{year}{2001}).

%
\end{thebibliography}
%


\newpage

%===============Figures


%\newpage
%===== TABLES

\end{document}